\documentclass[cits]{PoS}
\pdfoutput=1
\graphicspath{{img/}}
\usepackage[section]{placeins} % чтобы плавающие объекты не выходили за пределы своей секции

\title{Model of pp and AA collisions for the description of  long-range correlations}

\ShortTitle{Model of pp and AA collisions for the description of long-range correlations}

\author{\mauthor{Vladimir Kovalenko}%
										\\
        Saint Petersburg State University\\
        E-mail: \email{nvkinf@rambler.ru}}

\author{\speaker{Vladimir Vechernin}%
  										\\
        Saint Petersburg State University\\
        E-mail: \email{vechernin@gmail.com}}

\abstract{Soft processes in pp and AA interactions are considered in the framework of
phenomenological model with color strings formation and fusion. Elementary
parton collisions are realized in the model as the interaction of  two colour dipoles from
projectile and target nucleons. Modeling of the exclusive distributions of parton
momentum fractions and transverse coordinates is performed. The interaction of
colour strings in transverse plane is carried out in the framework of local string
fusion model with the introduction of the lattice in the impact parameter plane and
taking into account the finite rapidity length of strings. The parameters were fixed
with experimental data on pp total inelastic cross section and charged
multiplicity.

The model was used for the calculation of long-range correlations between the
multiplicities (n) and the mean transverse momenta (pt) of charged particles. The
dependence of n-n, pt-n, pt-pt correlations on the width and position of the
backward and forward rapidity windows was studied.

Note that the model enables to describe the AA interactions without referring
to the Glauber picture based on the concept of elementary nucleon-nucleon
collisions. In this connection the charged multiplicity, the mean numbers of
participant nucleons and binary collisions and their variances
in the case of PbPb collisions were calculated
and compared with the predictions of alternative models and the experimental
data. The influence of different ways of centrality determination on the multiplicity
fluctuations and long-range correlations was also discussed.}

\FullConference{XXI International Baldin Seminar on High Energy Physics Problems \\
		 September 10-15, 2012\\		 
		 JINR, Dubna, Russia}

\begin{document}

\section{Introduction}

The present work is devoted to development of
Monte Carlo model for the
description of soft proton-proton and heavy
ion collisions at high energy. Due to
inapplicability of usual QCD methods
in this region, such studies are semi-phenomenological.
One of the models, which successfully describe
particle production in soft region, is so-called
quark-gluon string model \cite{Capella,Werner1}.
The model involves two-stage scenario
of particles production: at the first stage
extended in rapidity objects (quark-gluon
strings, or colour tubes) are stretched between the partons
of colliding hadrons, and they fragment into
observable particles at the second stage.

In order to study these extended in rapidity strings
 it was proposed to measure long-range correlations
 between observables from two rapidity windows,
 separated by a gap.
 
 As these objects are finite in transverse plane,
the overlapping of them and possible interactions
can lead to non-trivial effects. Such
interaction of the strings is considered
in so-called string fusion model \cite{Braun1992154, Braun:1991dg,Braun00p349}.
In particular, this model predicts
the existence of non-zero long range correlations
between multiplicity and mean transverse momentum
(pt-n) and pt-pt correlations.
 The theoretical study of 
long-range correlations in the framework
of string fusion model is carried out
using Monte Carlo simulations \cite{VecherninYad,VecherninYadPt,Lakomov-en,Kovalenko10p257,KovalenkoYad}.
The goal of this work is to develop detailed
Monte Carlo model, taking into account
finite rapidity width of the strings and their fusion
at the transverse plane both for pp and AA collisions.

%One more \cite{DELPHICollaboration91p185}. qqq.

%About importance of sea quarks \cite{Kovalenko10p257}

%About importance of SF \cite{braun11,Braun:2003fn}

\section{The model}

In order to provide parton-string
model, which is suitable for reliable description of
fluctuations and correlations, one should
take into account not only inclusive,
but also exclusive parton distributions
as well as the details of the string formation and fusion.

\subsection{Partonic picture of p-p collision}

The form of inclusive distributions on the momentum fraction for N=2n partons is taken from \cite{7pdf,6pdf}:
		\begin{eqnarray}\label{structfFunctions}
\nonumber	   f_u(x)&=&f_{\bar u}(x)=C_{u,n}\ x^{-\frac{1}{2}}(1-x)^{\frac{1}{2}+n},
\\\nonumber	   f_d(x)&=&f_{\bar d}(x)=C_{d,n}\ x^{-\frac{1}{2}}(1-x)^{\frac{3}{2}+n},			   
\\\nonumber		f_{ud}(x)&=&C_{ud,n}\ x^{\frac{3}{2}}(1-x)^{-\frac{3}{2}+n},
\\\nonumber		f_{uu}(x)&=&C_{uu,n}\ x^{\frac{5}{2}}(1-x)^{-\frac{3}{2}+n}.
		\end{eqnarray}	
At $n>1$ the sea quarks and antiquarks have the same  distribution as the valence quarks.
Poisson distribution for the number of quark-antiquark 
(diquark) pairs ($n$) is assumed with some parameter $\lambda$, omiting the case of $n=0$,

For corresponding exclusive distribution for an arbitrary number of quark-antiquark pairs we find: 			\begin{eqnarray}\label{vyr}
			\rho(x_1,... x_N)=c\cdot\prod\limits_{j=1}^{N-1} x_j^{-\frac{1}{2}} 
							\cdot x_N^{\alpha_N} \delta(\sum\limits_{i=1}^N x_i -1).
			\end{eqnarray}

The valence quark is labelled by N-1, the diquark -- by N, and the other refers to sea quarks and antiquarks.			
${\alpha_N}$ = 3/2 (ud-diquark), ${\alpha_N}$ = 5/2 (uu-diquark).

Corresponding algorithms for modelling of such distributions are discussed in \cite{KovalenkoYad}.

The exclusive distribution in the impact parameter plane is constructed on the following suppositions: 
		\begin{enumerate}
			\item The position of the mass centre is fixed: $\sum\limits_{j=1}^N \vec r_j \cdot x_j =0$.
			\item The inclusive distribution of each parton is the \mbox{2-dimensional} Gaussian distribution.  
			\item The normalization condition is $<r^2>=<\frac{1}{N}\sum\limits_{j=1}^N {r_j}^2>={r_0}^2$,
		\end{enumerate}
where the parameter ${r_0}^2$ is connected with the mean square radius of the proton: \mbox{ $<r_N^2>=\frac{3}{2} r_0^2$.}

For the description of inelastic proton-proton collisions we assume that the elementary parton collision is implemented as an interaction of two color dipoles consisting of a valence quark and diquark, or of a quark-antiquark pair.
The probability amplitude of interaction of two dipoles with coordinates $(\vec r_1 \vec r_2)$ and $(\vec r_3 \vec r_4)$ 
in the impact parameter plane is given by \cite{Lonnbland1}: 
\begin{equation} \label{withlog}
	f=\frac{\alpha_S^2}{8}\ln^2 \frac {(\vec r_1 -\vec r_3 )^2 (\vec r_2 -\vec r_4 )^2 }
									{(\vec r_1 -\vec r_4 )^2 (\vec r_2 -\vec r_3 )^2 },
\end{equation}
where $a_S$ -- is a constant.

The value of $\alpha_S$ is assumed to be an effective coupling constant,
and $\alpha_S$ is used as a fitting parameter for better 
description of the experimental data, it is assumed that the value does not depend neither on the energy,
nor on the number of quark-antiquark pairs in the proton.

Taking into account the confinement effects \cite{Lonnbland1,Gustafson} leads:
		\begin{equation} \label{newformula}
		\nonumber
			f=\frac{\alpha_s^2}{2}\Big[ K_0\left(\frac{|\vec r_1-\vec r_3|}{r_{max}}\right) +
			K_0\left(\frac{|\vec r_2-\vec r_4|}{r_{max}}\right) 
			- K_0\left(\frac{|\vec r_1-\vec r_4|}{r_{max}}\right)
			- K_0\left(\frac{|\vec r_2-\vec r_3|}{r_{max}}\right)	\Big]^2
		\end{equation}		
which provides an exponential decrease of collision probability of two protons with large impact parameter.

In the eikonal approximation \cite{Gustafson,GustafsonNew}
 the probability of interaction of two dipoles is given by:
\begin{equation}\label{pij}
{p_{ij}=1-e^{-f_{ij}}}.
\end{equation}
The total probability of inelastic interaction of two ptorons is ${p=1-e^{-\sum\limits_{i,j} f_{ij}}}$,
where the summation is made over all the dipoles.

For more detiled discussion please refer also to \cite{Kovalenko10p257, KovalenkoYad}.

\subsection{Calculation of observables}

Probabilities of dipole interaction, obtained in the previous section,
 are used for the construction of the collision matrix. It should be
noted, that a dipole can interact only with one other dipole,
 so filling of the interaction matrix starts from the valence dipole.

The next step is the generation of strings in rapidity space.
Rapidity ends of a string $y_{min}, y_{max}$ are determined from the kinematic condition
of a string decay only on two particles with an average $p_t = 0.3 GeV$
and masses $m_{\pi}=0.15  GeV$ (for pion -- in case of quarl-antiquatk string) or $m_{p}=0.94  GeV$ (for nucleon - in case of diquark at the end of the string).

The strings, that are too short, are excluded from consideration by the requirement that the
sum of the masses of the particles produced should be less than the mass of the string,
 ie  $\sqrt{s} x_A x_B$, where $x_A, x_B$ --
  momentum fractions of the partons at the ends of the string. 
  Transverse coordinates of the center of the strings are equal to the arithmetic mean
of the corresponding coordinates of the partons at the ends.

Due to finite transverse size of the strings they overlap in the impact parameter plane. 
The interaction of colour strings in transverse plane is carried out in the framework of local string fusion model \cite{Braun00p349}
with the introduction of the lattice in the impact parameter plane. \cite{VecherninYad, VecherninYadPt, diskr2, Braun:2003fn, VecherninPoS}.
Cellular variant of string fusion has been chosen in order to simplify taking into account finite rapidity
length of strings, because, as it had been shown, numerical results
differ little in several string fusion model variants \cite{VecherninYad,Lakomov-en}.

According to this model, mean multiplicity of charged particles and mean $p_t$ originated from the area $S_k$, where k strings are overlapping are the following:
\begin{equation} \label{muptloc}\nonumber
	\left\langle \mu\right\rangle_k=\mu_0 \sqrt{k} \frac{S_k}{\sigma_0} \hspace*{1cm}
	\left\langle p_t\right\rangle_k=p_0 \sqrt[4]{k},
\end{equation}	 
where $\sigma_0=\pi r_{str}^2$ -- string transverse area

In the discrete model transverse plane is considered as a grid with the cell area equals to string transverse area and strings are fused if their centers occupy the same cell.

Finite rapidity width of strings is also taken into account\cite{KovalenkoYad}:
rapidity bins are defined by strings edges, so every rapidity interval contains constant number of strings,
then bins are processed separately with the final summation over whole rapidity axis.

The Monte Carlo model was implemented as C++ class,
and in order to get results with high statistics,
distributed grid computing have been involved \cite{KovalenkoGrid}.

\subsection*{Parameters fixation}

In the framework of our model the growth of multiplicity and total inelastic cross section is achieved by increasing the number of quark-antiquark pairs (parameter $\lambda $).
The value of $\lambda $ is turned from the experimental data on the total inelastic cross section \cite{sigmafit}.

\begin{figure}[!htbp]
\center
\includegraphics[width=.4\textwidth]{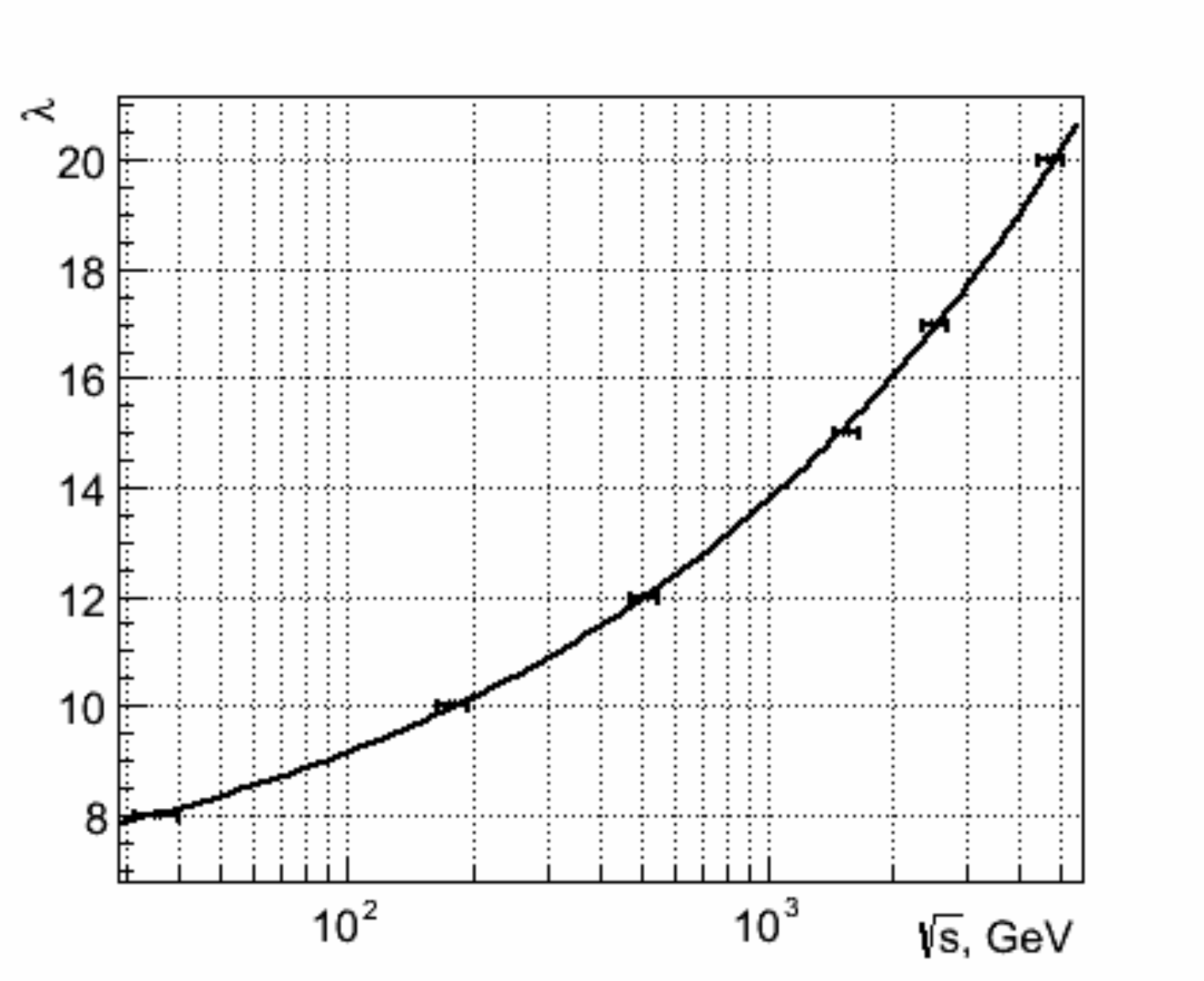}
\caption{Connection of parameter $\lambda$ and the energy for the case of $\alpha_s = 0.7$.
}
\label{lamote}
\end{figure}

The simulations are conducted in two phases.
The first step is devoted to determination of 
$\sigma^{in}_ {pp}$ for the given $\lambda$. 
Using experimental data, the dependence of $\lambda$
on collision energy is obtained
 (example is shown at fig. \ref{lamote}).

One should note the importance of sea quark presence
even at low energy \cite{Kovalenko10p257}.

\FloatBarrier

At the second step all other calculations
are carried out for the given energy.

The value of mean multiplicity per rapidity from one single string $\mu_0$ is fixed at one point at low energy.

The other parameters of the model were chosen as follows:
     for $ r_{max} \simeq 0.2-0.3 $ fm; ratio $ r_{max} / r_0 $ is chosen to be $ 0.5 $;
     constant $\alpha_S$ is rixed for best description of the dependence of charged multiplicity on energy \cite{Aamodt:2010pp}.
     For the string radius $ r_{str} $ we
considered several possible values:
     $ 0.2fm, 0.3fm, 0.4fm $, which are within the estimates \cite {braunp, diskr1, diskr2, Braun:2003fn},
used for the description of long-range correlations in nucleus-nucleus collisions.

\begin{figure}[!htbp]
\begin{center}
\includegraphics[width=.65\textwidth]{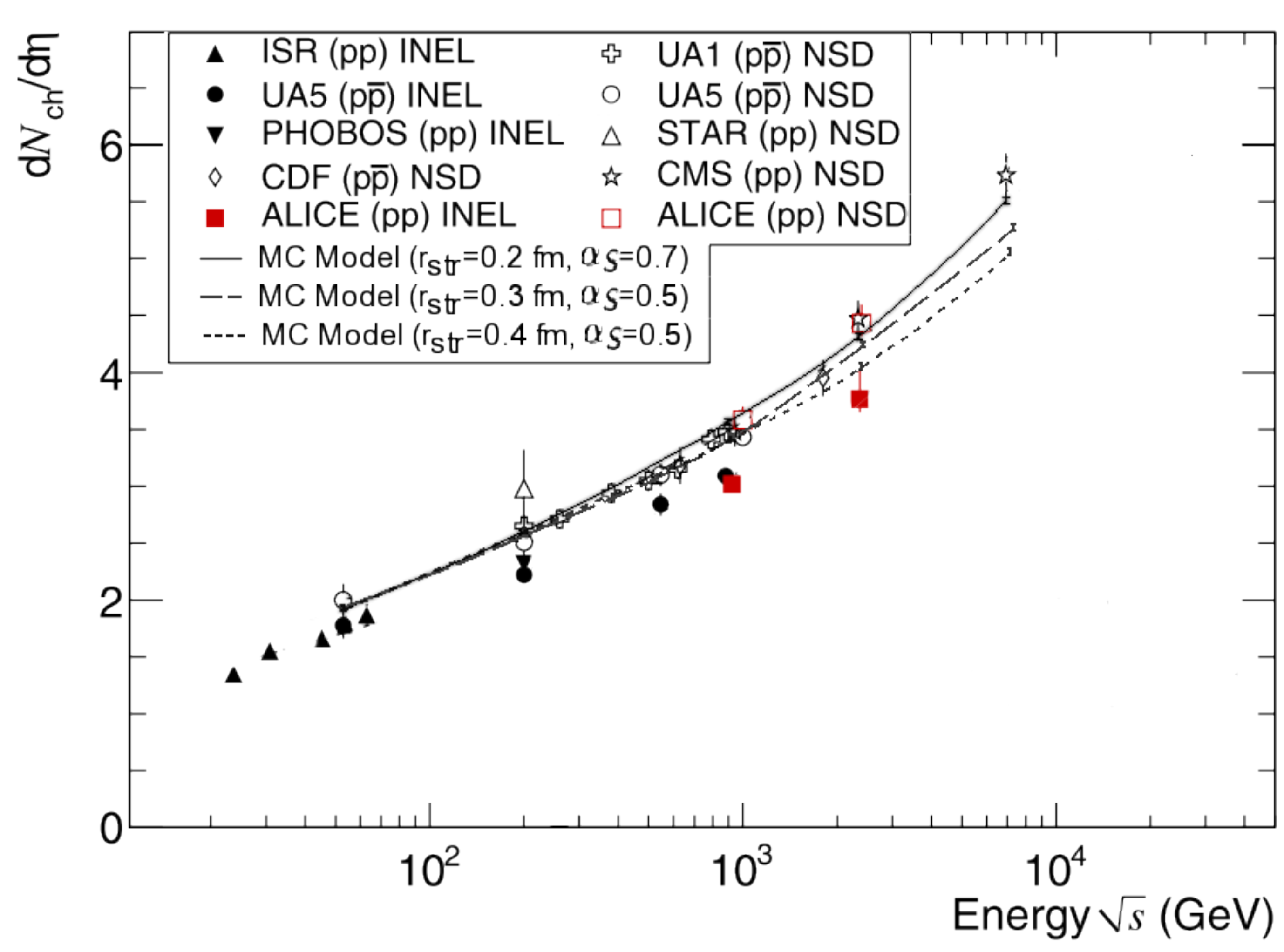}
\caption{Chaged-particle multiplicity in p-p collisions.
Calculations of the Monte Carlo model and experimental data
\cite{Aamodt:2010pp,Khachatryan:2010us}.
}
\label{multv}
\end{center}
\end{figure}

The results of parameters fixing are presented in Table \ref{tabl1},
the corresponding plot for the multiplicity shown in
Fig. \ref{multv}.

\begin{table}[!h]
\caption{}
\label{tabl1}
\begin{center}
\begin{tabular}{|l|l|l|l|}
\hline
$r_{str}$=0.2fm & $\alpha_S$=0.7 &  $\mu_0$=0.92  \\ \hline
$r_{str}$=0.3fm & $\alpha_S$=0.5 &  $\mu_0$=1.02  \\ \hline
$r_{str}$=0.4fm & $\alpha_S$=0.5 &  $\mu_0$=1.12  \\ 
\hline
\end{tabular}
\end{center}
\end{table}\FloatBarrier

\FloatBarrier
The calculation results show good description of the multiplicity in a wide energy range, and taking
into account the results of the LHC at 7 TeV
we use the first set of parameters ($ r_{str} $ = 0.2fm, $\alpha_S$ = 0.7, $\mu_0 $ = 0.92) for further studies.

%jkkjkj
%\begin{figure}[!htbp]
%\begin{center}
%\includegraphics[width=.6\textwidth]{sigmatotinel}
%\caption{This is the caption of the figure.
%\cite{Aamodt:2010pp,Khachatryan:2010us}
%}
%\label{sigmatotinel}
%\end{center}
%\end{figure}\FloatBarrier

\subsection*{Calculation of correlation functions}

The correlation function
between two observable B (backward) and F (forward)
from different rapidity windows is defined
as the mean value of B with fixed F \cite{VecherninYad}:

\begin{equation}
f_{B-F}(F)={\langle B\rangle}_F.
\end{equation}

The correlation coefficient represents the slope
of correlation function:
\begin{equation} \label{defcorr}
b_{B-F}=\frac{df(F)}{dF}|_{{F}=<{F}>}.
\end{equation}

Often it is useful to switch to
normalized variables: $B \rightarrow B/{\langle B\rangle}, F \rightarrow F/{\langle F\rangle}$, in this case
 $p_t-n$ correlation coefficient would become dimensionless,
 and both $n-n$ and $p_t-p_t$ correlation coefficients
 would not change in case of symmetrical windows.
Thus we use the following definitions:
\begin{eqnarray}
b_{nn}^{}&=&\frac{<n_F>}{<n_B>} \cdot\frac{d<{n}_B>}{dn_F}|_{n_{F}=<n_{F}>}, \\
b_{p_t-n}^{}&=&\frac{<{n}_F>}{<{p_t}_B>} \cdot\frac{d<{p_t}_B>} {d{n}_F}|_{n_{F}=<n_{F}>}, \\
b_{p_t-p_t}^{}&=&\frac{<{p_t}_F>}{<{p_t}_B>} \cdot\frac{d<{{p_t}}_B>}{d{p_t}_F}|_{{p_t}_{F}=<{p_t}_{F}>}.
\end{eqnarray}

In practice, the experimental determination of the correlation coefficient and Monte Carlo simulation is performed by obtaining fitting of $f(F)$ by linear function from ${\langle F\rangle}-\sigma_{{\langle F\rangle}}$ to ${\langle F\rangle}+\sigma_{{\langle F\rangle}}$,
where $\sigma_{{\langle F\rangle}}=\sqrt{<F^2>-{\langle F\rangle}^2}$. This method is used in this paper.

For the calculation of these correlations in the framework
of string fusion model
 the occupancy  $\eta_i$ of each cell
is defined from string configuration and one assumes, that
every cell of the lattice emit particles according to
Poisson distribution with parameter equal to $\eta_i \cdot \mu_0 \Delta y$. The mean event transverse momentum
is generated with Gaussian distribution with mean \cite{VecherninYadPt}
$
(\sum\limits_i n_i \alpha_i)/(\sum\limits_i n_i)
$
and variance $\sum\limits_i n_i D_i$,
where $\alpha_i=\sqrt[4]{p_0},$\  $D_i=\sqrt{\eta_i}D,$ \  $D=\gamma p_0$.

The coefficient $\gamma$ is the coefficient of proportionality between the average transverse momentum and root of the variance of
transverse momentum from one single string.
This parameter is assumed to be independent of the energy.
As in \cite{VecherninYadPt}, we used the value $\gamma = 0.61 $.

Obtained by this way mean event transverse
momentum and multiplicity in both forward and backward
windows are used to will 2d histogram (fig. \ref{nncloud2}) and correlation function is obtained by regression.

This way in case of n-n and pt-n correlations is fully equivalent to the one, described in \cite{VecherninYad,KovalenkoYad} and also coincides with
the one, used for pt-pt correlations \cite{VecherninYadPt}.

\section{Results on pp interaction}

\subsection*{Correlation functions}

\begin{figure}[!htbp]\center
\includegraphics[width=.65\textwidth]{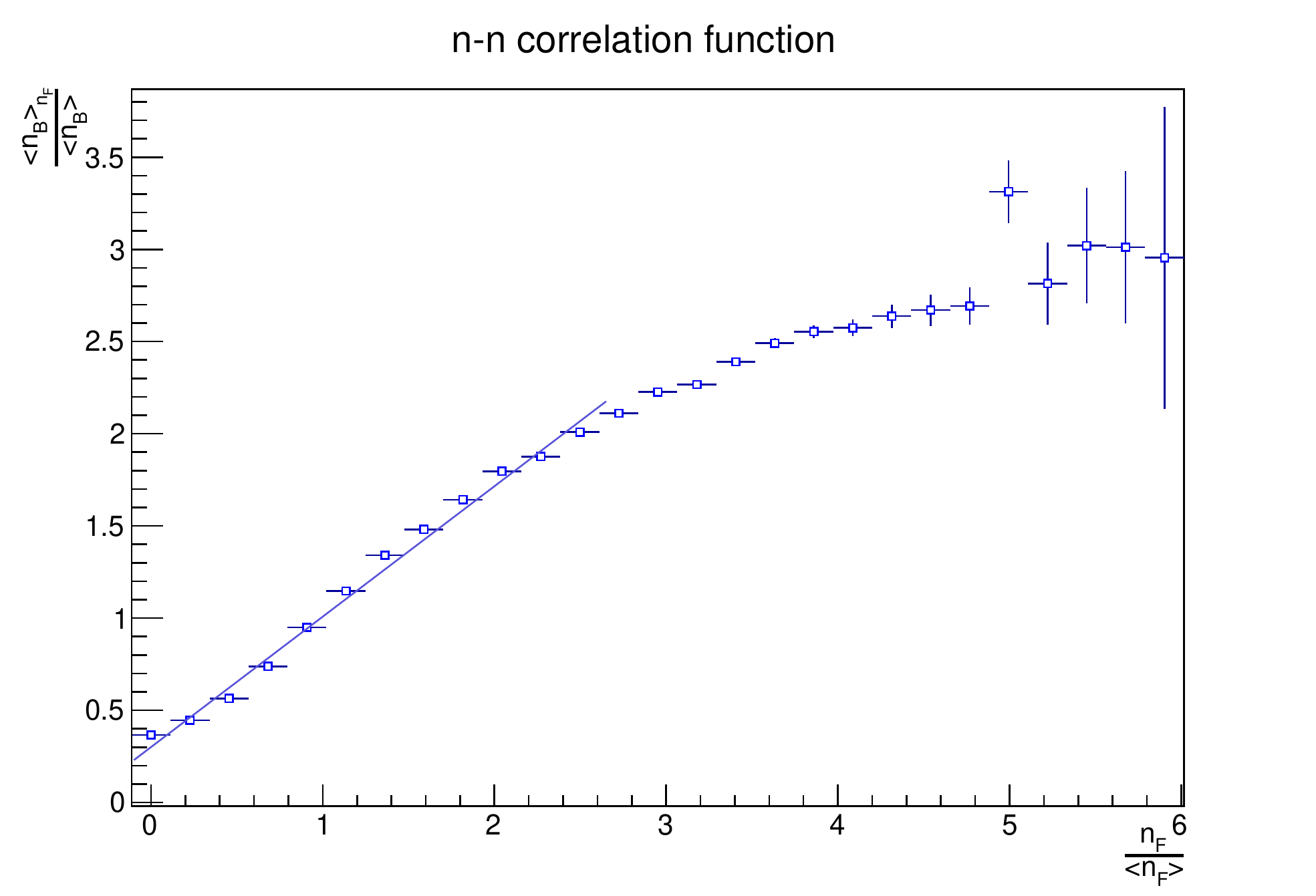}
\caption{n-n correlation function for pp collisions at 7 TeV, calculated in MC model. Rapidity windows are (-0.8, 0) and (0 ,0.8).
}
\label{nn7000}
\end{figure}

\begin{figure}[!htbp]\center
\includegraphics[width=.65\textwidth]{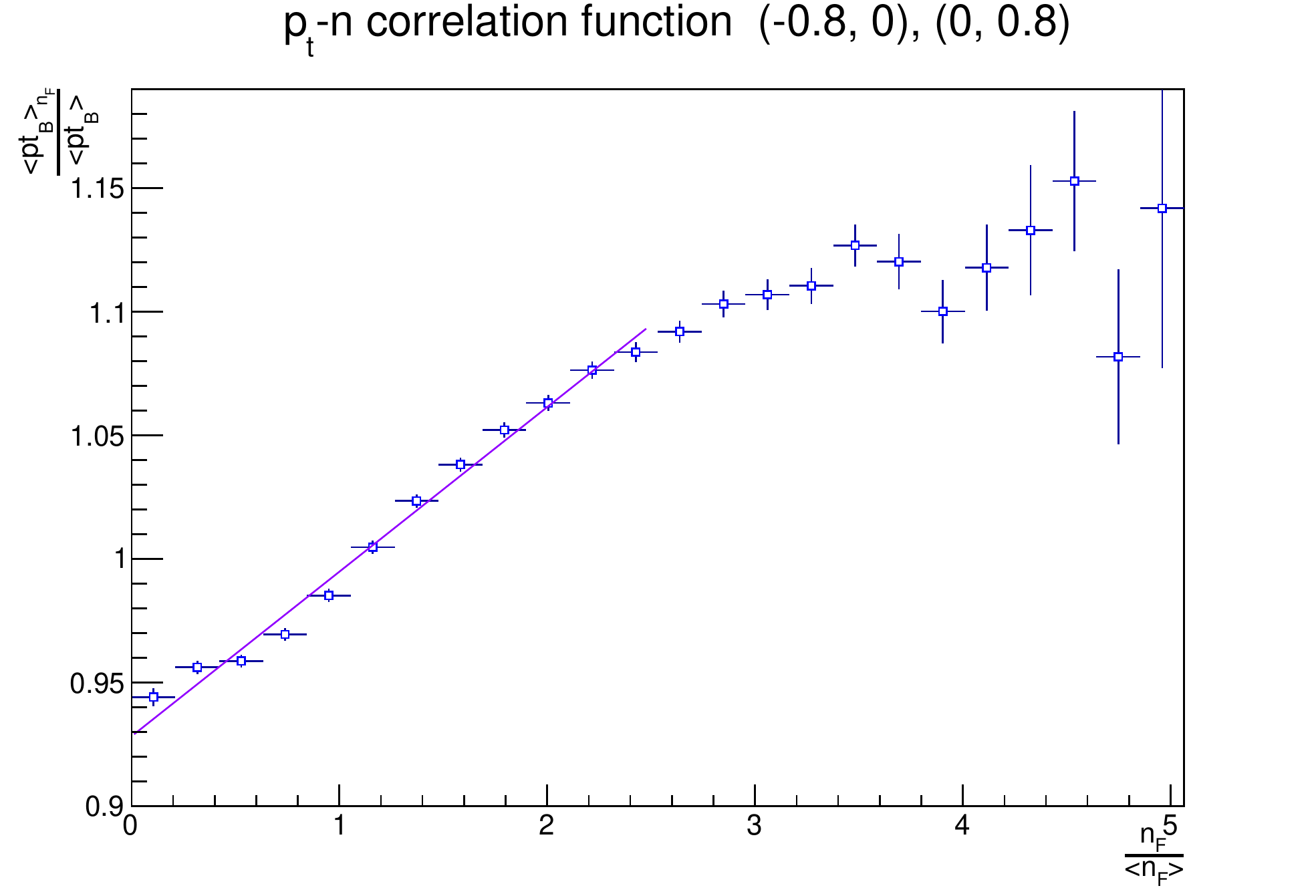}
\caption{pt-n correlation functions for pp collisions at 7 TeV, calculated in MC model.
}
\label{ptn7000}
\end{figure}

\begin{figure}[!htbp]\center
\includegraphics[width=.65\textwidth]{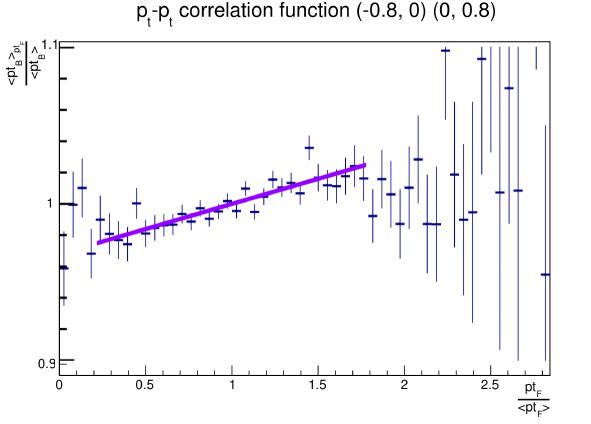}
\caption{pt-pt correlation functions for pp collisions at 7 TeV, calculated in MC model.
}
\label{ptpt7000}
\end{figure}

Correlation functions for p-p collisions are sown at fig. \ref{nn7000}, \ref{ptn7000} and \ref{ptpt7000}. One should note the non-linearity of all types correlation functions with significant deviation from linear function occuring only outside the region $ \langle n_F \rangle - \sigma_F < n_F <  \langle n_F \rangle + \sigma_F $, which is used for determination of correlation coefficients.

\FloatBarrier

\FloatBarrier

\FloatBarrier

\subsection* {Correlation coefficients}

We also studied the dependence of correlation coefficients
on the width of rapidity windows.

\begin{figure}[!htbp]\center
\includegraphics[width=.6\textwidth]{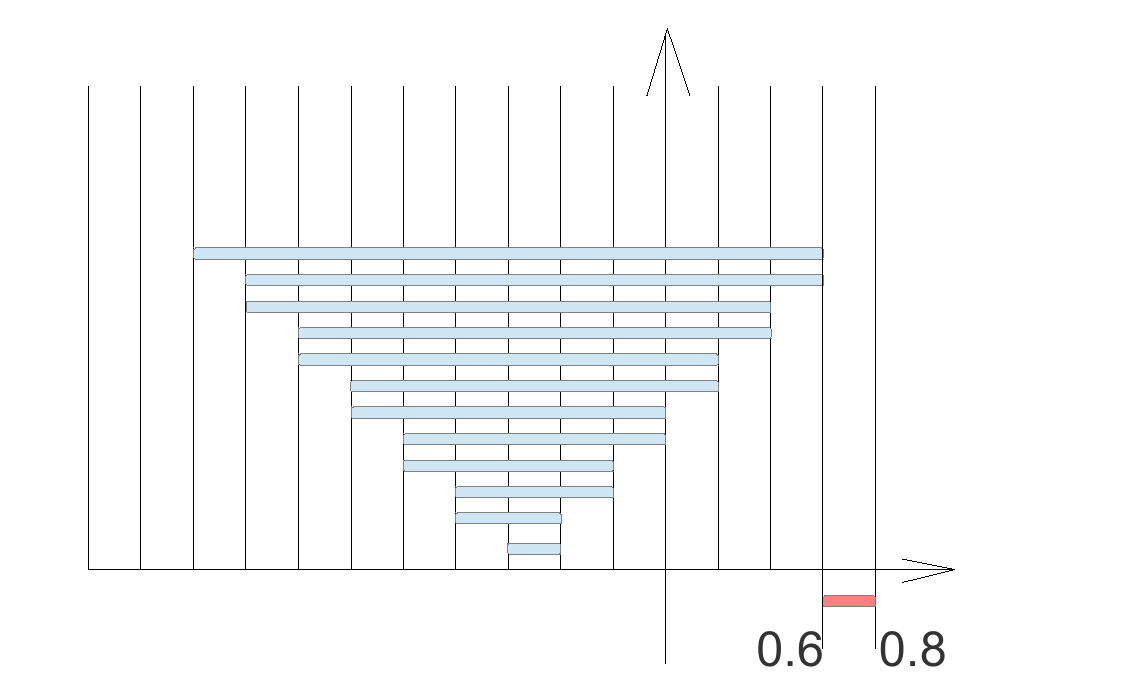}
\caption{Configuration of rapidity windows with fixed forward window.
}
\label{shemab}
\end{figure}

\begin{figure}[!htbp]\center
\includegraphics[width=.42\textwidth]{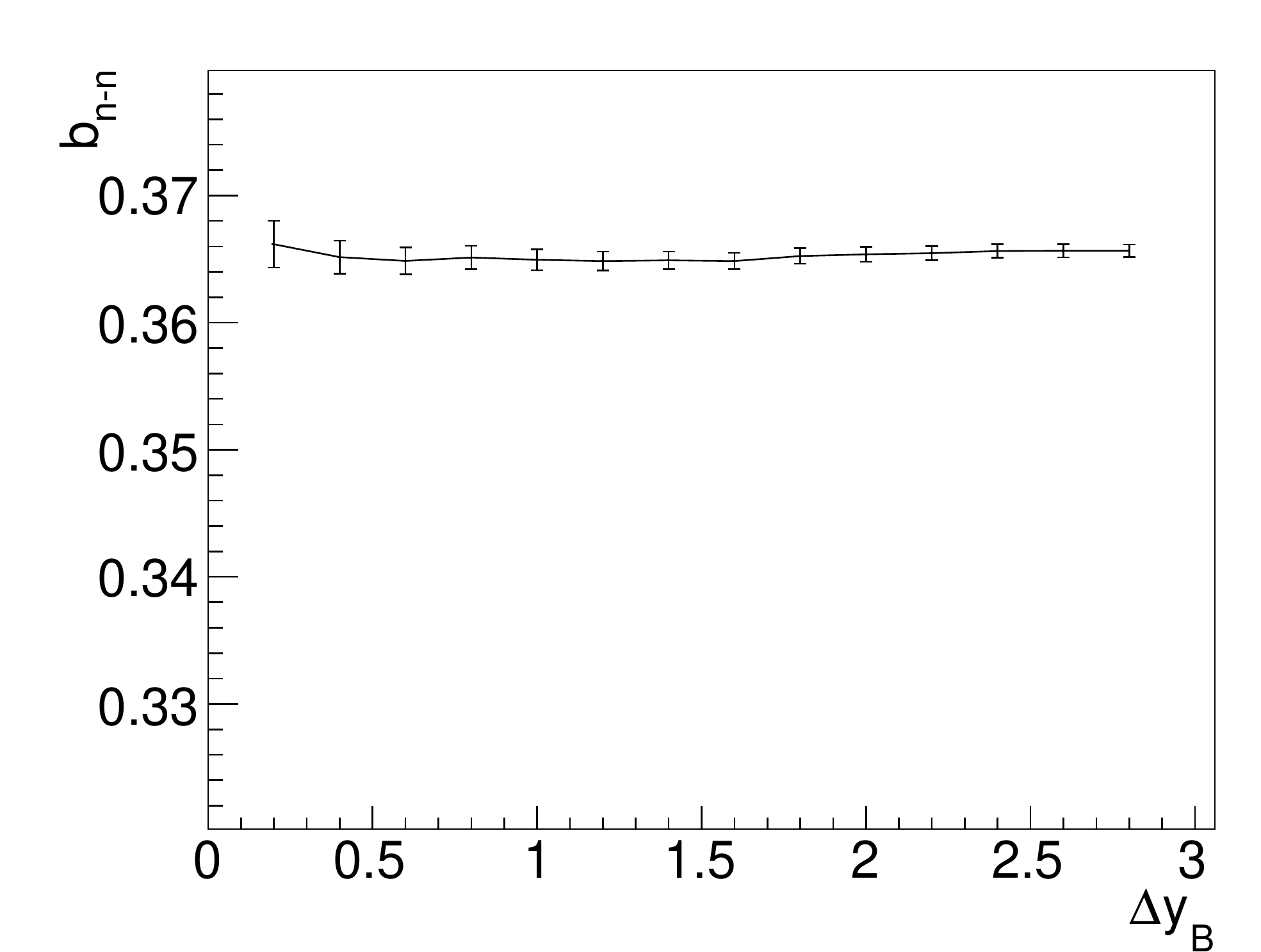}

\includegraphics[width=.42\textwidth]{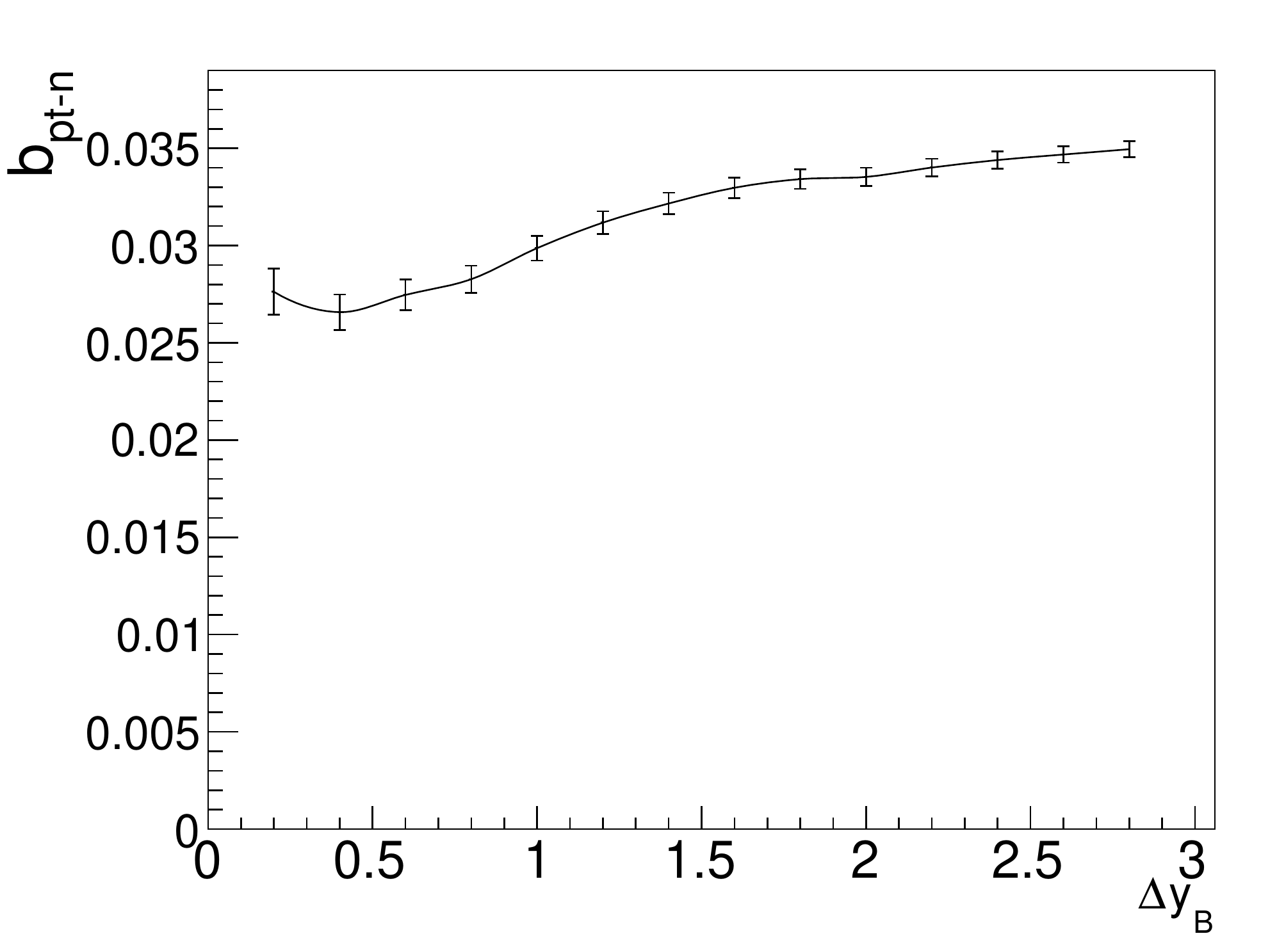}
\includegraphics[width=.42\textwidth]{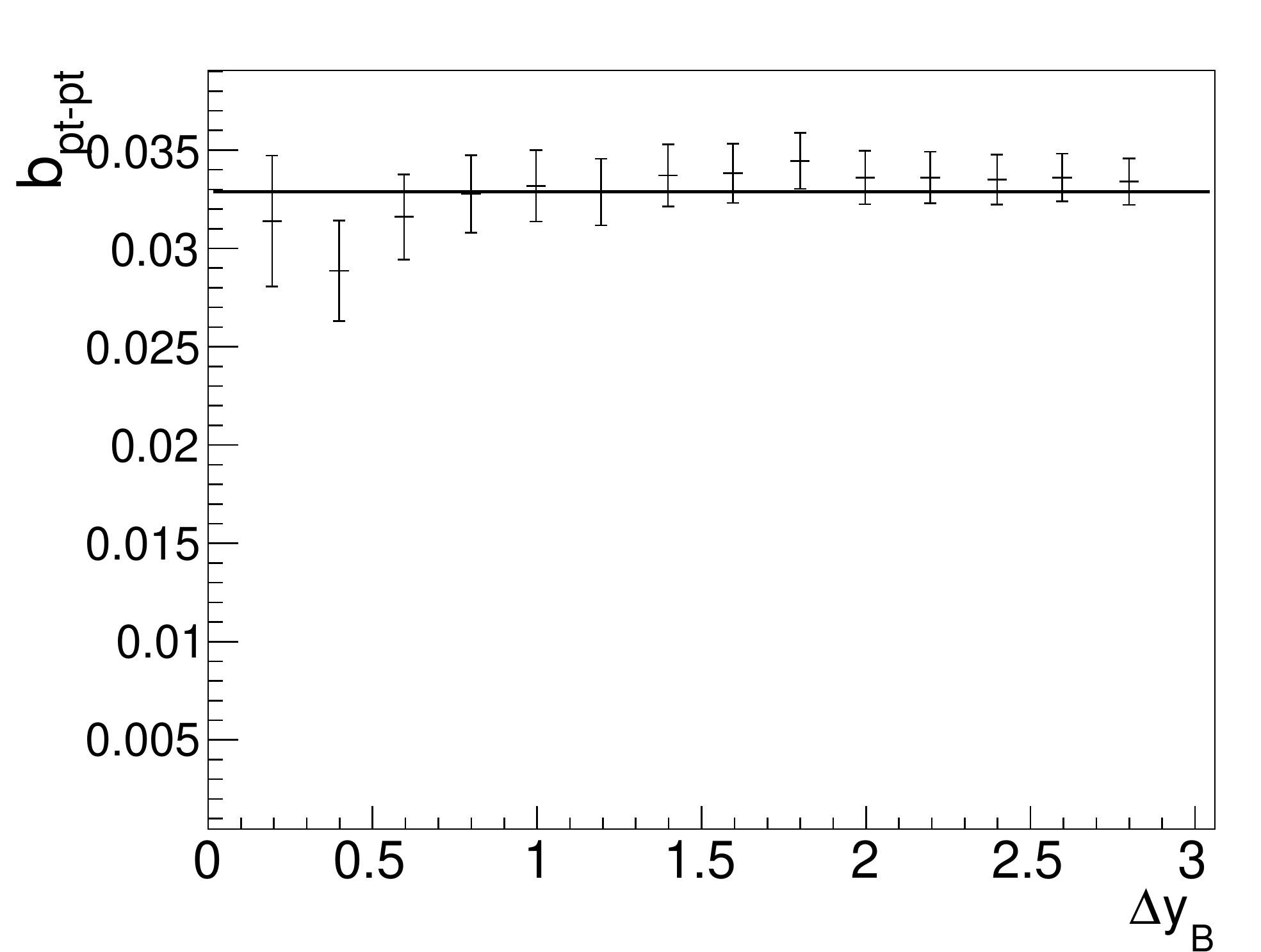}
\caption{n-n, pt-n and pt-pt correlation coefficients
in p-p collisions at 7 TeV as a function of the width of backward window.
}
\label{bdyb}
\end{figure}

Firstly we consider the dependence of correlation coefficients on the width of \textit {backward} rapidity window with fixed forward window at (0.6, 0.8). The configurations of the windows used are shown at the fig. \ref {shemab}.
Correlation coefficients are calculated for pp collisions at 7 TeV and results are shown at the fig. \ref{bdyb}.

\FloatBarrier

\FloatBarrier

\begin{figure}[!htbp]\center
\includegraphics[width=.42\textwidth]{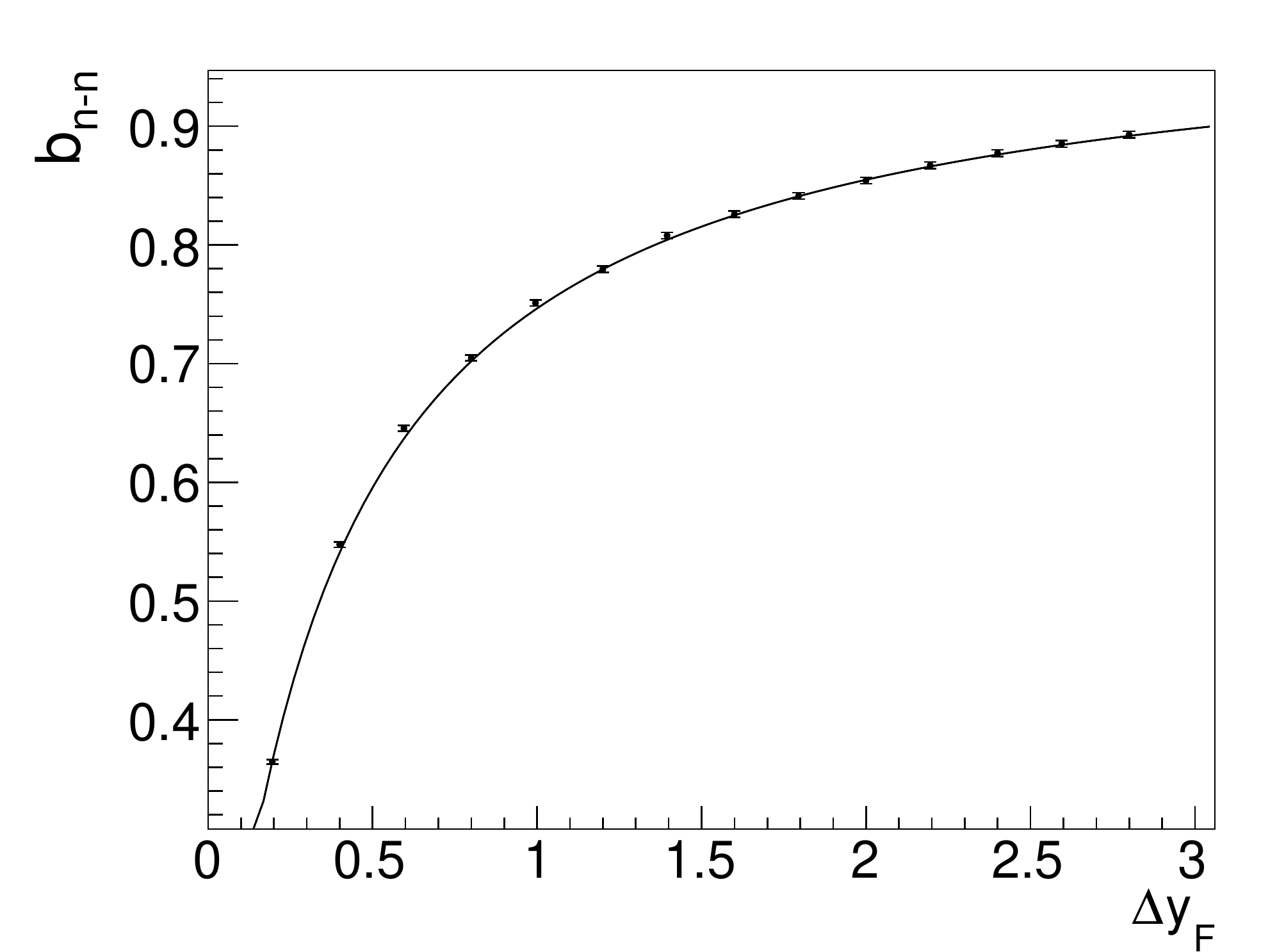}

\includegraphics[width=.42\textwidth]{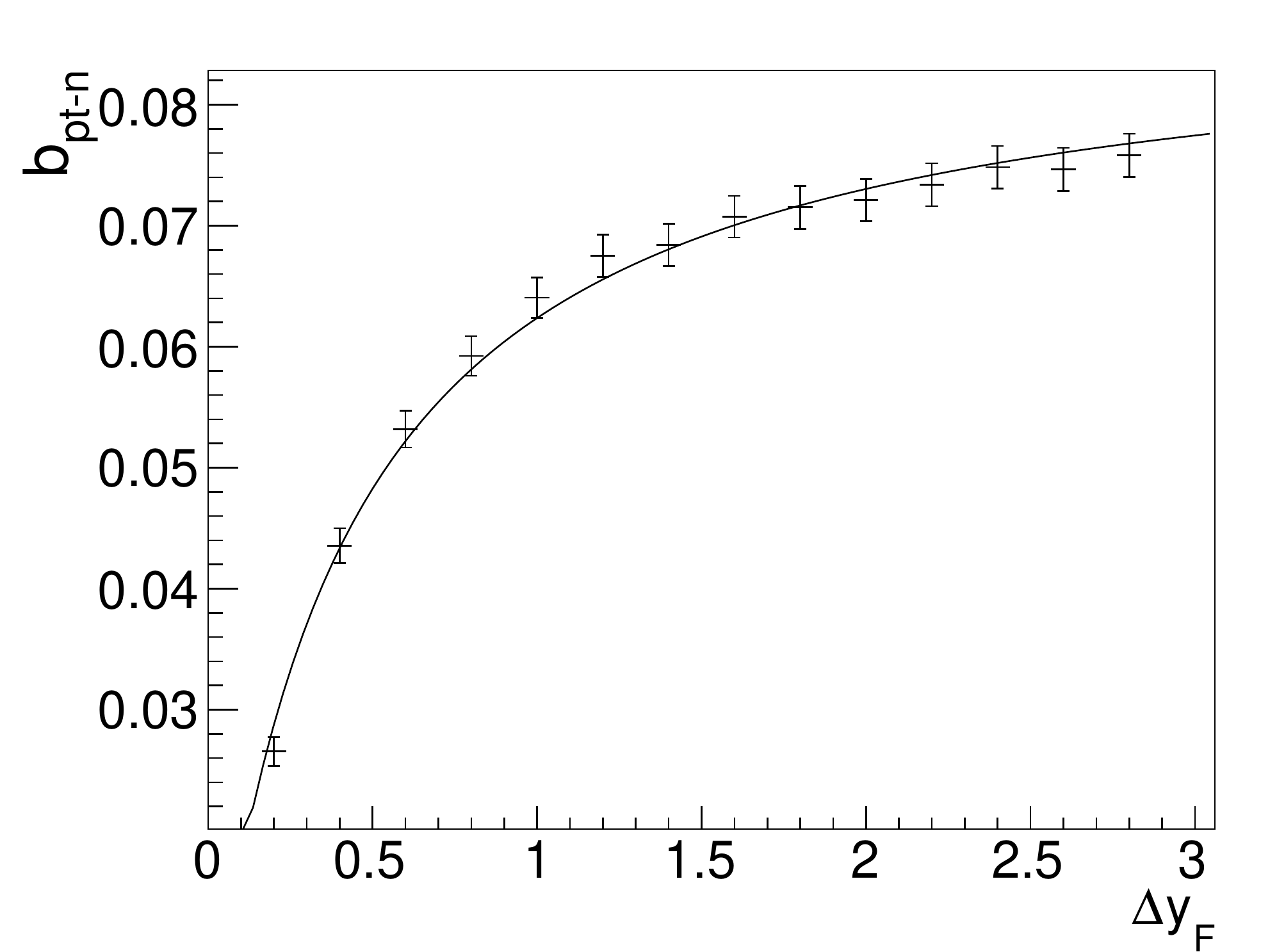}
\includegraphics[width=.42\textwidth]{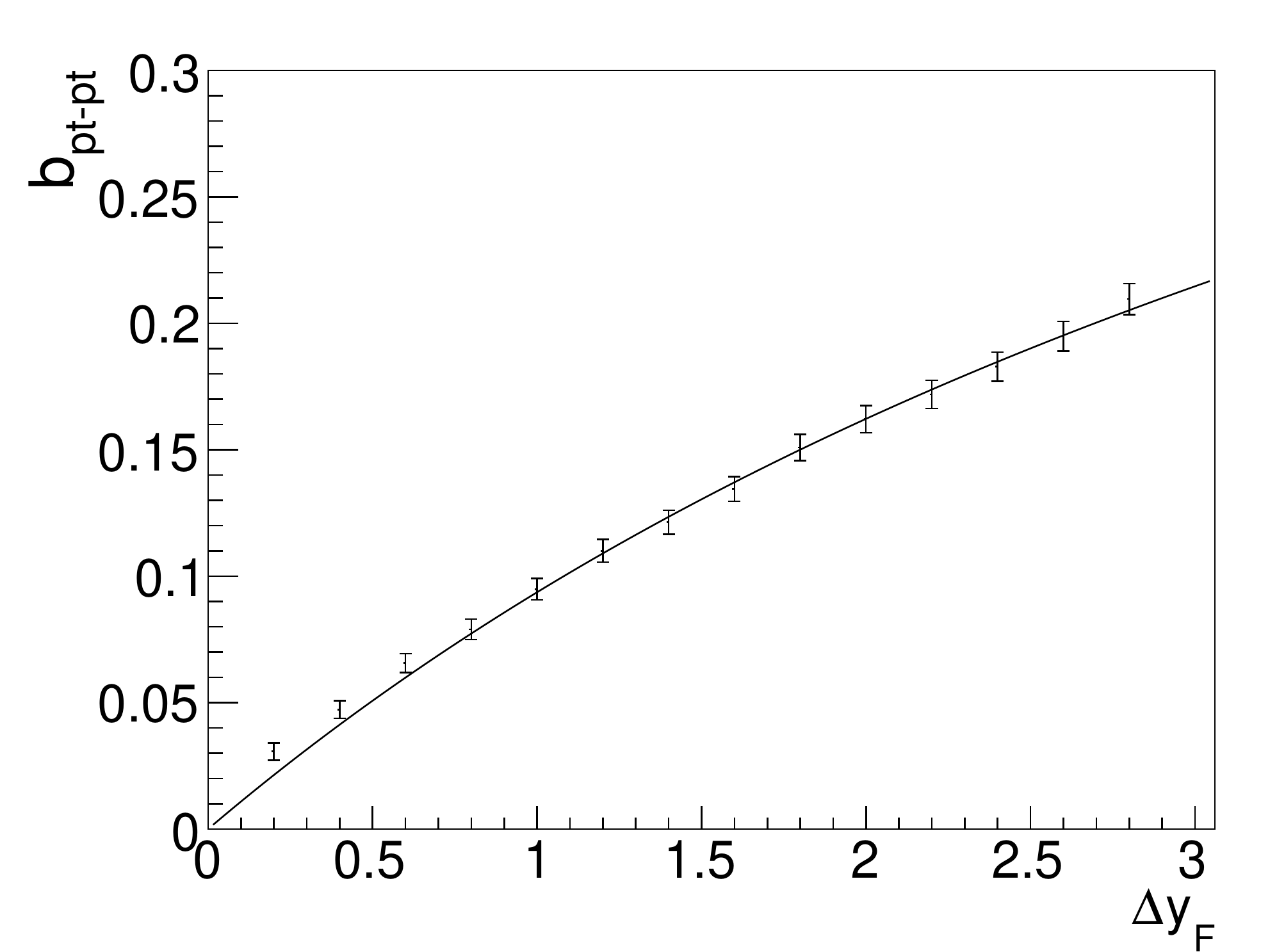}
\caption{n-n, pt-n and pt-pt correlation coefficients
in p-p collisions at 7 TeV as a function of the width of forward window.
}
\label{bdyf}
\end{figure}

The dependence of correlation coefficients on the width of \textit{forward} rapidity windows is shown at fig. \ref{bdyf}.

\begin{figure}[!htbp]\center
\includegraphics[width=.6\textwidth]{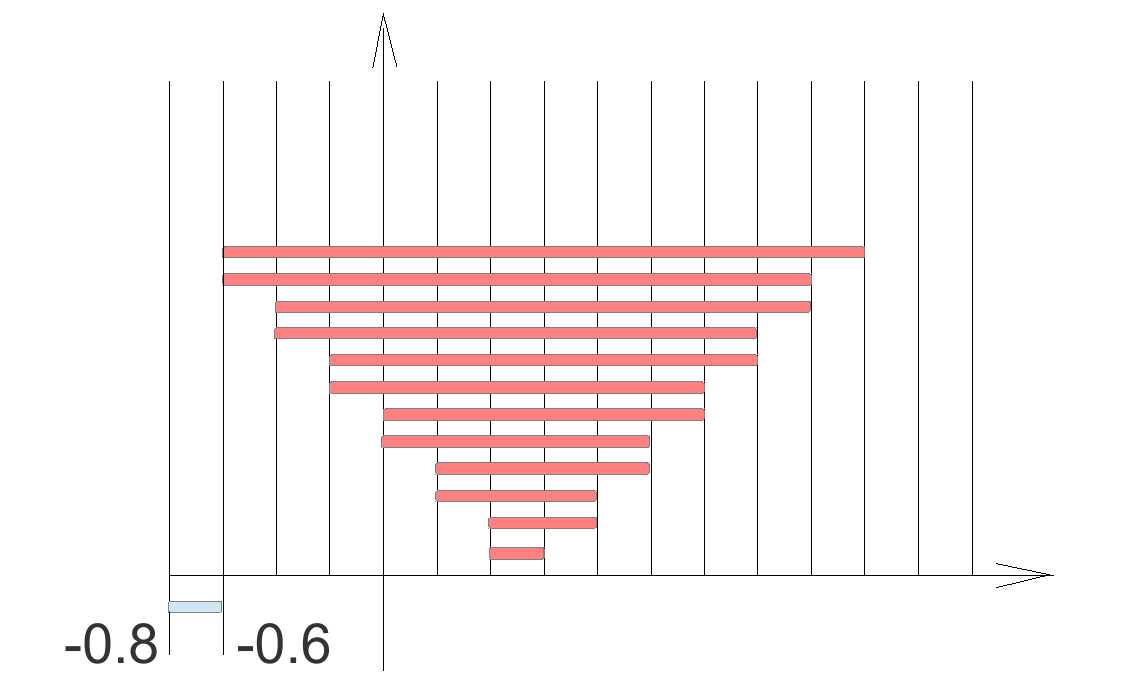}
\caption{Configurations of rapidity windows with fixed backward window.
}
\label{shemaf}
\end{figure}\FloatBarrier

\FloatBarrier

The values calculated in the Monte Carlo program
are fitted by the formula
\begin{equation}\label{nnfa2}
b =\beta \frac {\Delta y_F} {\Delta y_F+k}.
\end{equation}
Fit parameters, obtained for n-n, pt-n and pt-pt correlations are shown in the table \ref {tabl2}.

\begin{table}%[H]
\caption {Parameters of a fitirovaniye of coefficients of correlation}
\label {tabl2}
\begin {center}
\begin {tabular} {|c|c|c|}
\hline
Type of correlation & $\beta$ & $k$ \\ \hline
$n-n$ & 1.000 $\pm$0.002 & 0.34 $\pm$0.01 \\ \hline
$p_t-n$ & 0.088 $\pm$0.002 & 0.41 $\pm$0.02 \\ \hline
$p_t-p_t$ & 0.53 $\pm$0.05 & 4.6 $\pm$0.6 \\ \hline
\end {tabular}
\end {center}
\end {table}\FloatBarrier

\subsubsection* {Discussion of results}

The absence of dependence of $n-n$ correlation coefficient
on the width of a backward window corresponds to the predictions
of the models of independent emitters \cite{Vechernin:2010tg} and to asimptotical
expressions, obtained in the string fusion model \cite{VecherninYad,VecherninYadPt, Kolevatov:2005ey} in the limit case of high string density.

Small dependence of $p_t-n$ correlation coefficient
on the width of a forward window can be
 caused by dependence of this coefficient
on the gap between windows (which is stronger for $p_t-n$ correlation, than for $n-n$). An additional source of
such dependence may be constraint $n_B>0$. In fact, if we select an event with non-zero number of particles in backward window, 
we potentially select higher-multiplicity event, probably, with higher mean transverse momentum and
this can change the value of the correlation
coefficient in the relative variables.
In wide backward window this constraint becomes negligible.

$P_t-p_t$ correlations
also do not depend on width of a back window within error bars.

The dependence of a coefficient of correlation on the
width of a forward window is completely described by the formula (\ref {nnfa2}); for $n-n$ correlation $\beta=1$
is found, that is in full agreement with the predictions
 of the model of the independent emitters \cite{Vechernin:2010tg}.
The fact, that the value $k$ for $n-n$ and $p_t-n$ correlations  does not coincide, as it was
predicted for high string density \cite{VecherninYad}, means that the density of strings, where the asymptotic formula  is applicable,  is not reached in p-p collisions at 7 TeV.

One should notice that $p_t-p_t$ correlations strongly depend on width of forward rapidity window, and at wide sufficiently $\Delta y_ {F} $ dominate over $p_t-n$ correlations.

From the physical point of view that fact that 
correlation coefficients depend only on the width of a forward window means \cite {diskr2,diskr1} that the dynamic variable in the forward window is the one that classifies events,
and the wider window is, the closer this classification coincides with the one by string configurations, and hence, the more this value is correlated with a variable in another rapidity window.

In addition, one should notice that obtained dependence of
correlation coefficients on the width of rapidity
windows can be confirmed in experiment; this is done for $n-n$ correlations \cite {FeofilovQM, FeofilovPoS}, and used for the efficiency corrections.

\

\

\

\pagebreak

\subsubsection* {Dependence of correlation coefficients on the gap between rapidity windows}

\begin{figure}[!htbp]\center
\includegraphics[width=.62\textwidth]{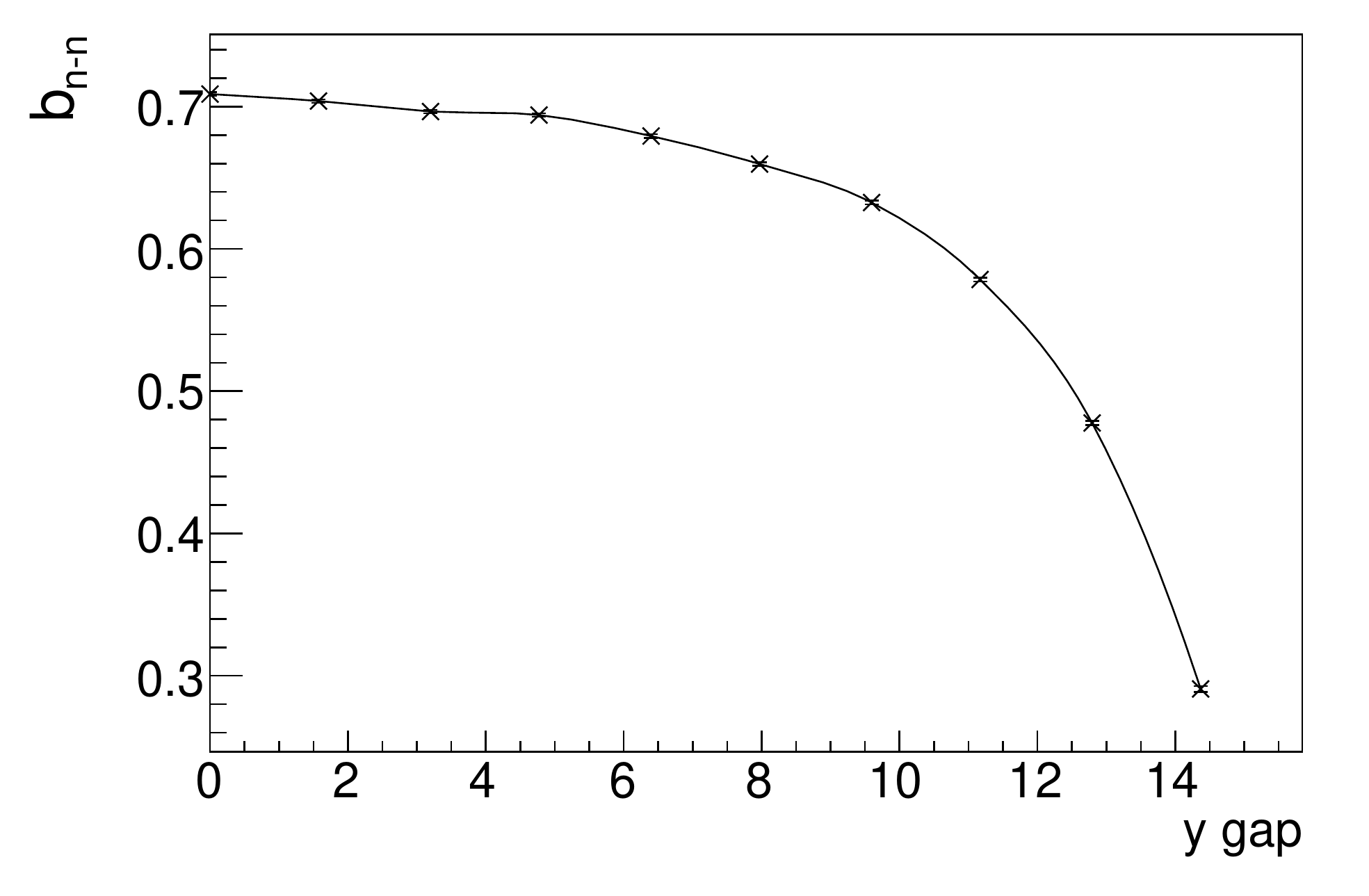}
\includegraphics[width=.62\textwidth]{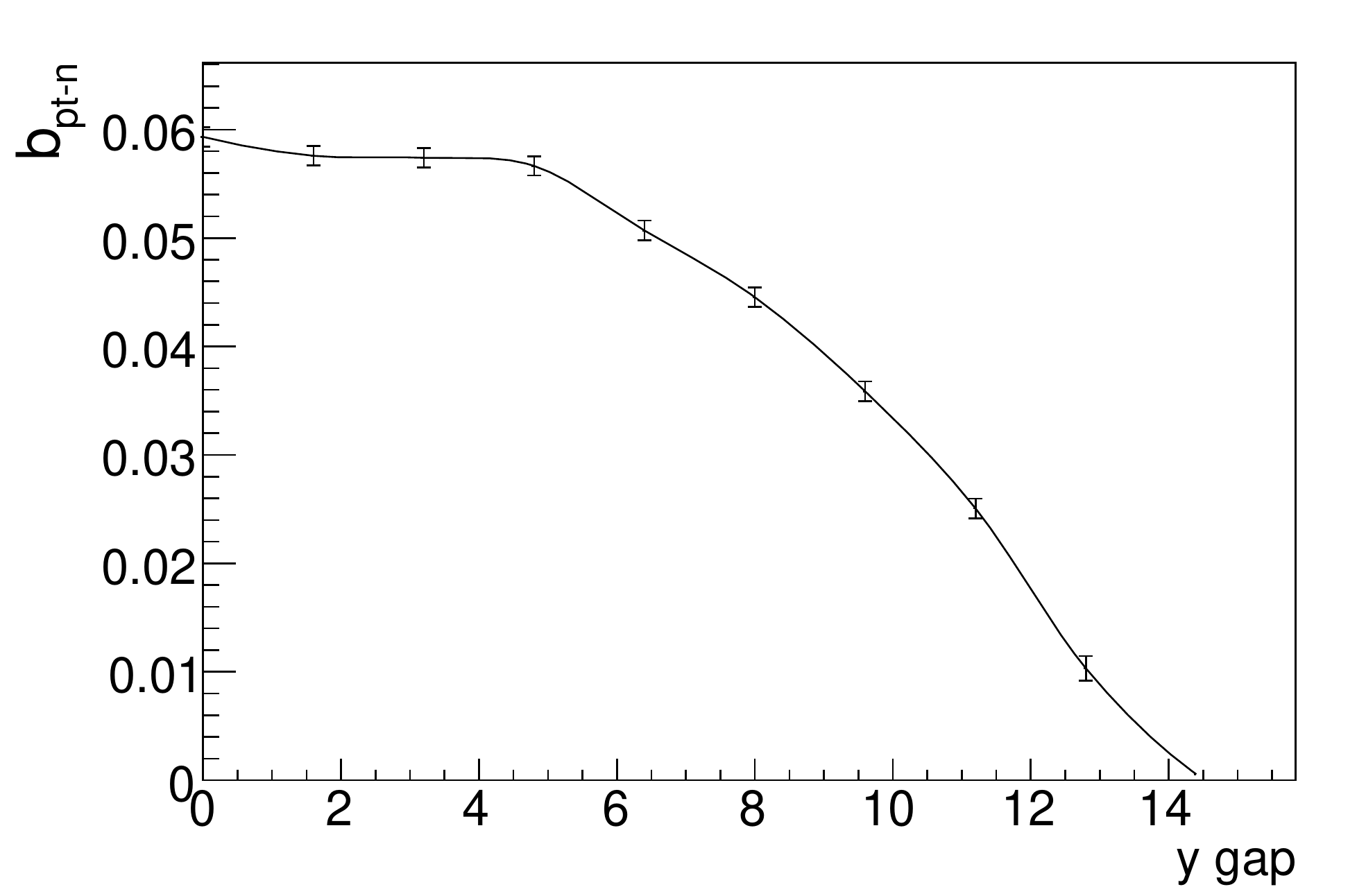}
\includegraphics[width=.62\textwidth]{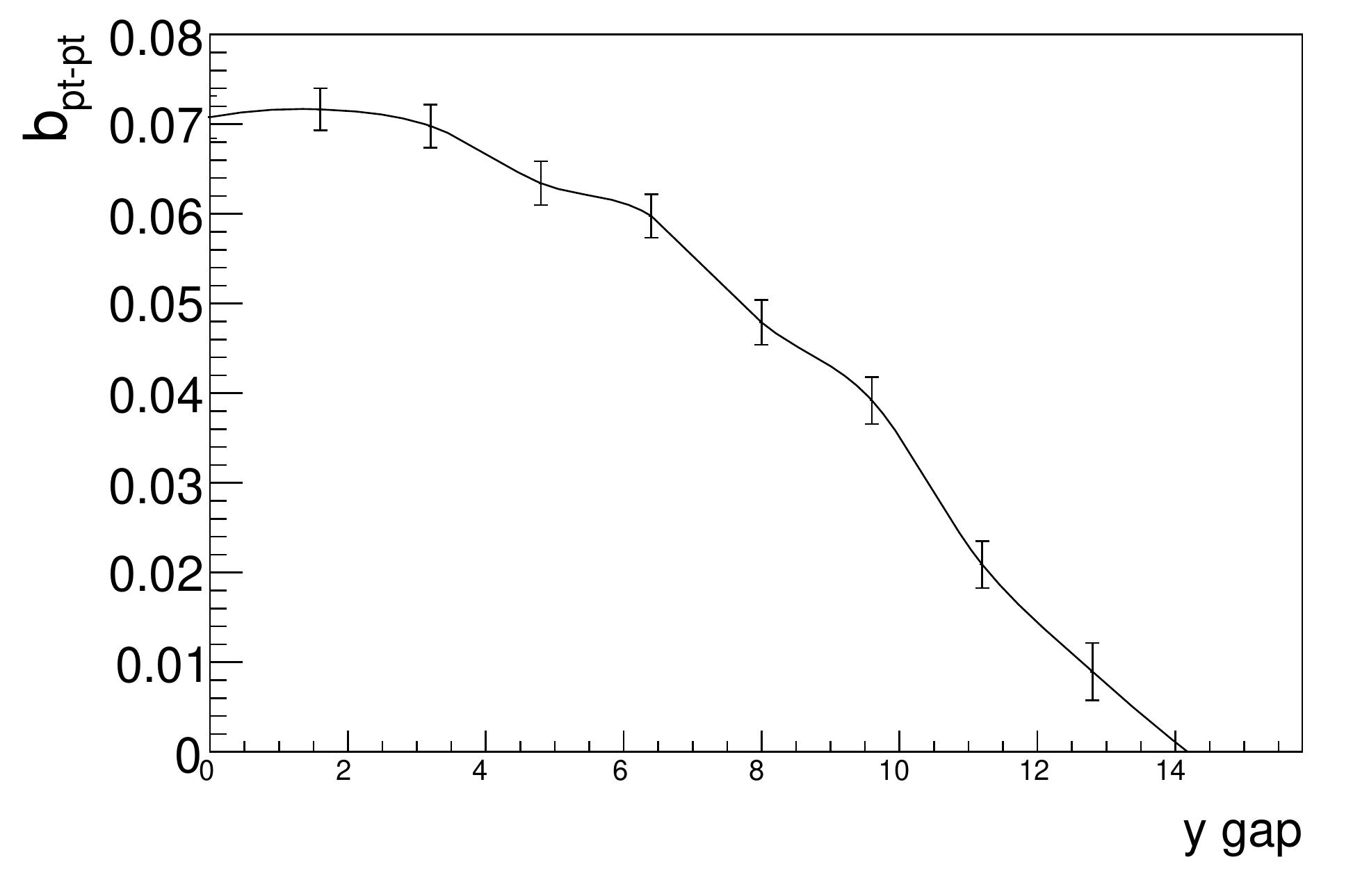}
\caption{Dependence of n-n, pt-n and pt-pt correlation coefficients on the gap between rapidity windows
}
\label{bgap}
\end{figure}

The dependence of correlation coefficients 
on the provision of rapidity windows
was also studied in the model. The symmetric windows of
the width 0.8 rapidity units were chosen.
The  results are shown at the fig. \ref{bgap}.

\FloatBarrier

All three types of correlations
show similar behavior and decrease
at high gap in the same way, as rapidity distribution of charged particles.

It is noticeable that both $p_t-n$ and $p_t-p_t$  
correlation coefficients reach zero at wide enough rapidity gap. It can be it is explained as follows:
only the lengthiest valence strings contribute
to these windows far away from midrapidity,
therefore there is no strings to fuse in this area, and consequently there are no $p_t-n$ and $p_t-p_t$ correlation.
Multiplicity-multiplicity correlations remain due to fluctuation of the
numbers of strings (volume fluctuation): zero and one.

%On fig. \ref {bA} coefficients of correlation are shown
%for pp of collisions at energy of 7 TeV
%in average area быстрот as functions
%from a gap between windows also it is made 
%comparison with experimental data.
%
%Somewhat it is possible to consider comparison as conditional, as
%\begin {enumerate}
%\item in model of a window get out on speed, on experiment
%on pseudo-speed;
%\item in experimental data is available trimming on cross impulse as the model is calculated on the description of all weak area up to zero for $p_t$;
%\item at data interpretation is used
%"full-range" a way of a fitirovaniye, which with the account
%forms of correlative function can yield results 
%below, than фит $\pm of \sigma$ used in this work.
%\end {enumerate}

%In addition, in case of a small gap are essential
%correlations bound to disintegrations of resonances, and it
%leads to the strong dependence from width of a gap on experiment.
%In favor of it says that fact that in a case 
%azimuthal splitting of windows on sector
%in the developed windows dependence on a gap
%practically there is no \cite {FeofilovQM}.

%Taking into account these reservations it is possible to argue that 
%calculations of model don't contradict the experimental
%to data.

%Thus, in this chapter were analysed
%correlative AND functions coefficients of correlation,
%in a proton - proton collisions,
%their dependence on energy, width and provisions
%bystrotny windows comparison of results also is executed with
%data of experiment of ALICE.

\section{Nucleus-nucleus collisions}

For the description of nucleus-nucleus collisions
at high energy the Glauber model is is widely used \cite{GlauberFull,UrsSelTop}, which is based 
on an assumption of incoherent superposition of
single nucleon-nucleon collisions.
It is supposed that trajectories of nucleons in a nuclei are approximately straight lines and all consequent
collisions are occurred at the same cross section and with identical mean number of produced charged particles (the same as for pp collision).

Thus, multiplicities in pp and AA collisions are connected by the following:
\begin {equation}\label{simplemult}
N_{ch}^{AA} = N_{col} N_{ch}^{pp}.
\end {equation}

In such manner Glauber model considerably overestimate charged multiplicity in comparison with experimental data and contains obvious breaking of the energy conservation:
indeed, the in such picture same part of energy can go to the particle production several times \cite{feofivanov}.

In order to get an agreement with experimental
data one should use instead of (\ref {simplemult}) some interpolations:
\begin {equation}
N_{ch} = C (x N_{part} + (1-x) N_{coll }).
\end {equation}
With turned parameters $C$ and $x$ one can achieve  of distribution of charged multiplicity consistent with experimental data. 
The Glauber model is used in such manner for the definition of number of participating nucleons in those experiments where it is impossible to determine it directly \cite{Aamodt:2010cz}.

This problem is considered also in modified Glauber model \cite{feofivanov}, where energy loss is taken into
account effectively. There are also several models \cite{PSM,Braun03p79, Amelin:1993cs} representing "The Glauber on a parton level".

In the present paper we develop a Monte Carlo
model without referring
to the Glauber picture based on the concept of elementary sequential nucleon-nucleon or partonic collisions.
We implement direct generalization of pp model, 
described above.

The initial arrangement of nucleons is done with standard Woods-Saxon distribution:
\newline
$\rho_A (r) = \frac {\rho_0} {1 + exp [(r - R) / d] }
$ with $R = 1.07 \cdot A^ {1/3} fm$, $d = 0.545fm$.

Nucleons are treated as set of the dipoles; the elementary collisions it is carried out by means of formulas (\ref {newformula} - \ref{pij}). The nucleon is considered 
as participant if at least one of its dipoles faced a dipole of other nucleus.
There is no additional parameter turning.
The set of observables is calculated in the same way as for proton-proton collisions. 

\begin{figure}[!htbp]\center
\includegraphics[width=1.02\textwidth]{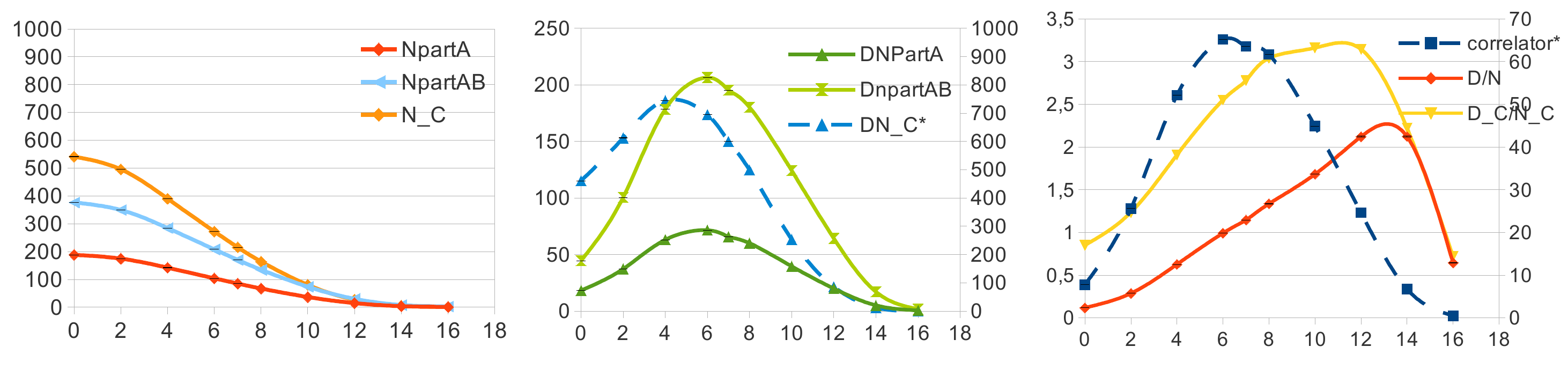}
\includegraphics[width=1.04\textwidth]{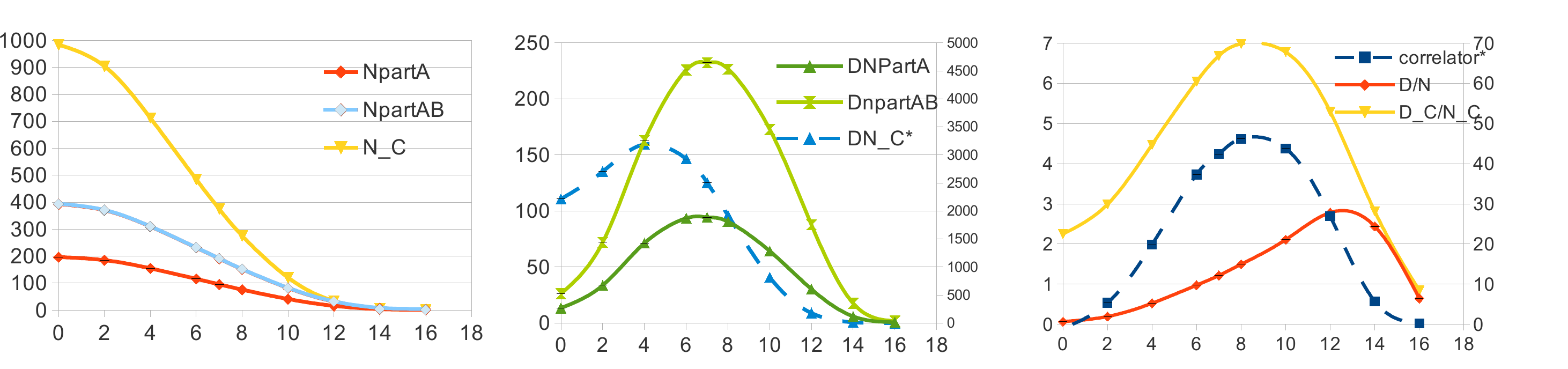}
\caption{Number of participant nucleons, number of binary collisions (left), their variances (middle), scaled variances and
correlator $2(\langle N_{part_A} N_{part_B}\rangle -
\langle N_{part_A}\rangle \langle N_{part_B}\rangle) $ (right):
our Monte-Carlo model without Glauber picture (up), and calculations in the Glauber model at $\sigma_{NN}=34mb$ (down).\newline
note: additional scale (at the right) is used for variables, marked by star.}
\label{models}
\end{figure}\FloatBarrier

In the framework of the MC model  we calculated the number of participants, binary collisions and their variances for
PbPb collisions at low energy ($\sigma_{NN}=34 mb$)
at fixed impact parameter
and compare them with results of Glauber model calculations (fig. \ref{models}).

The results show practically the complete coincidence of predictions of two models on the number of participated nucleons, while the number of binary collisions is almost twice less. There is qualitatively similar behavior of variances of number of participants and number of collisions in these two models; quantitatively in the Glauber model
fluctuations appear higher.

The calculations carried out by the author in the Glauber model are close with analytical and numerical calculations \cite {Vechernin07, Vechernin12} $\sigma_ {NN}  =31.4mb$.

The decrease of the number of binary collisions
can be explained by the following: in every nucleus-nucleus
collision each dipole can interact with other one only once, 
and the energy of colliding partons directly goes to
particle production. It reduces additional
probability of further interactions of the nucleons with
already "used" partons. Thus, such picture
respects the energy conservation.

\begin{figure}[!htbp]\center
\includegraphics[width=0.6\textwidth]{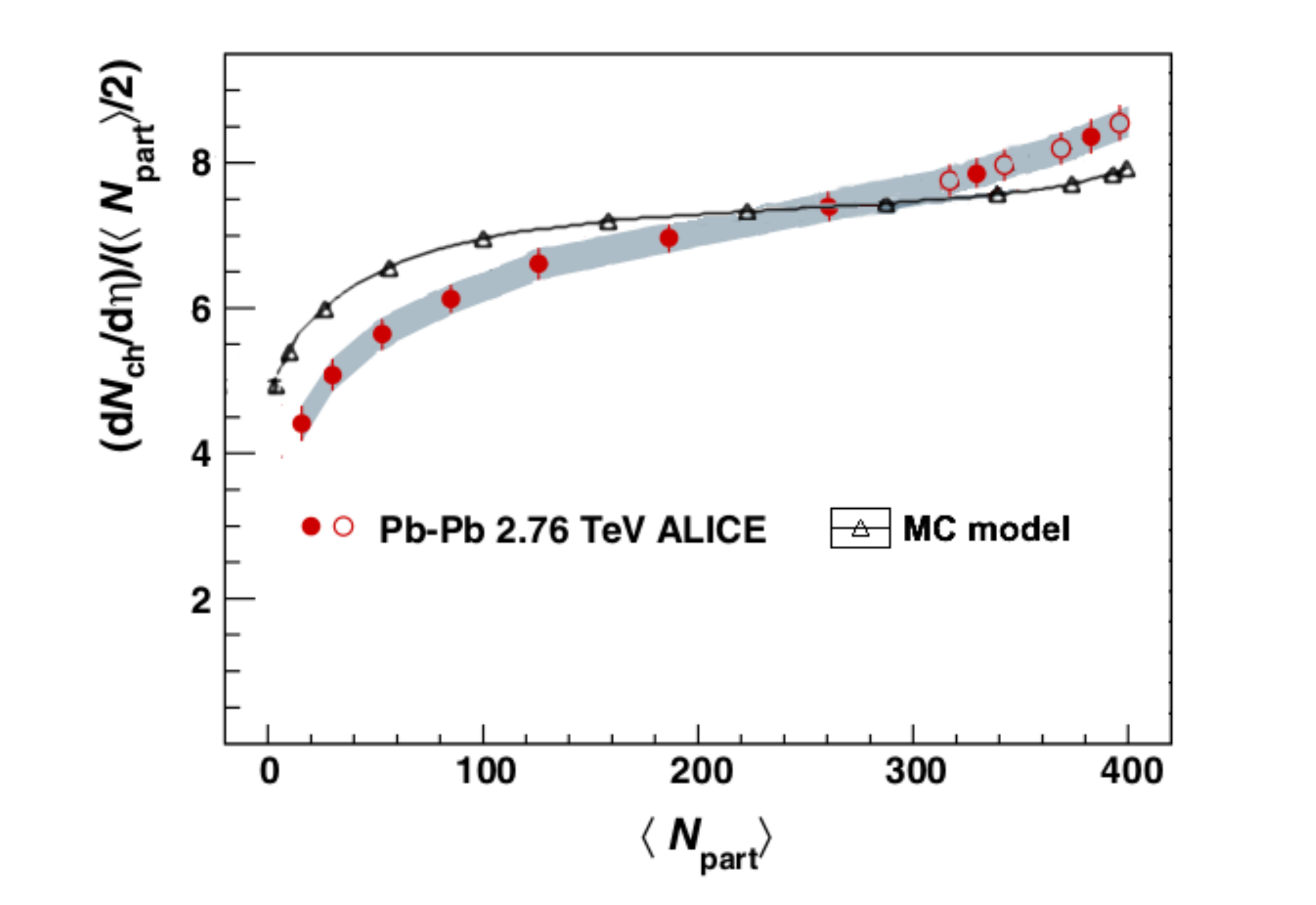}
\caption{Dependence of $(dN_{ch}/d\eta)/(N_{part}/2)$ on the number of participants for Pb-Pb collisions at $\sqrt{s_{NN}}=2.76$TeV.
\  Results of MC calculations and experimental data \cite{Aamodt:2010cz}.
}
\label{dndnpart}
\end{figure}

The charged multiplicity
per rapidity over 
 a half of number of participating nucleons 
at the LHC energy
is shown at the fig. \ref{dndnpart} and
compared with ALICE experimental data.

In general, we achieved a good agreement, remaining
discrepancy probably could be reduced by
additional parameters turning, taking into account
some results on p-A or AA scattering. 
%HIJING event generator \cite{Wang91p3501, Aamodt:2010cz}

\FloatBarrier

\begin{figure}[!htbp]\center
\includegraphics[width=.7\textwidth]{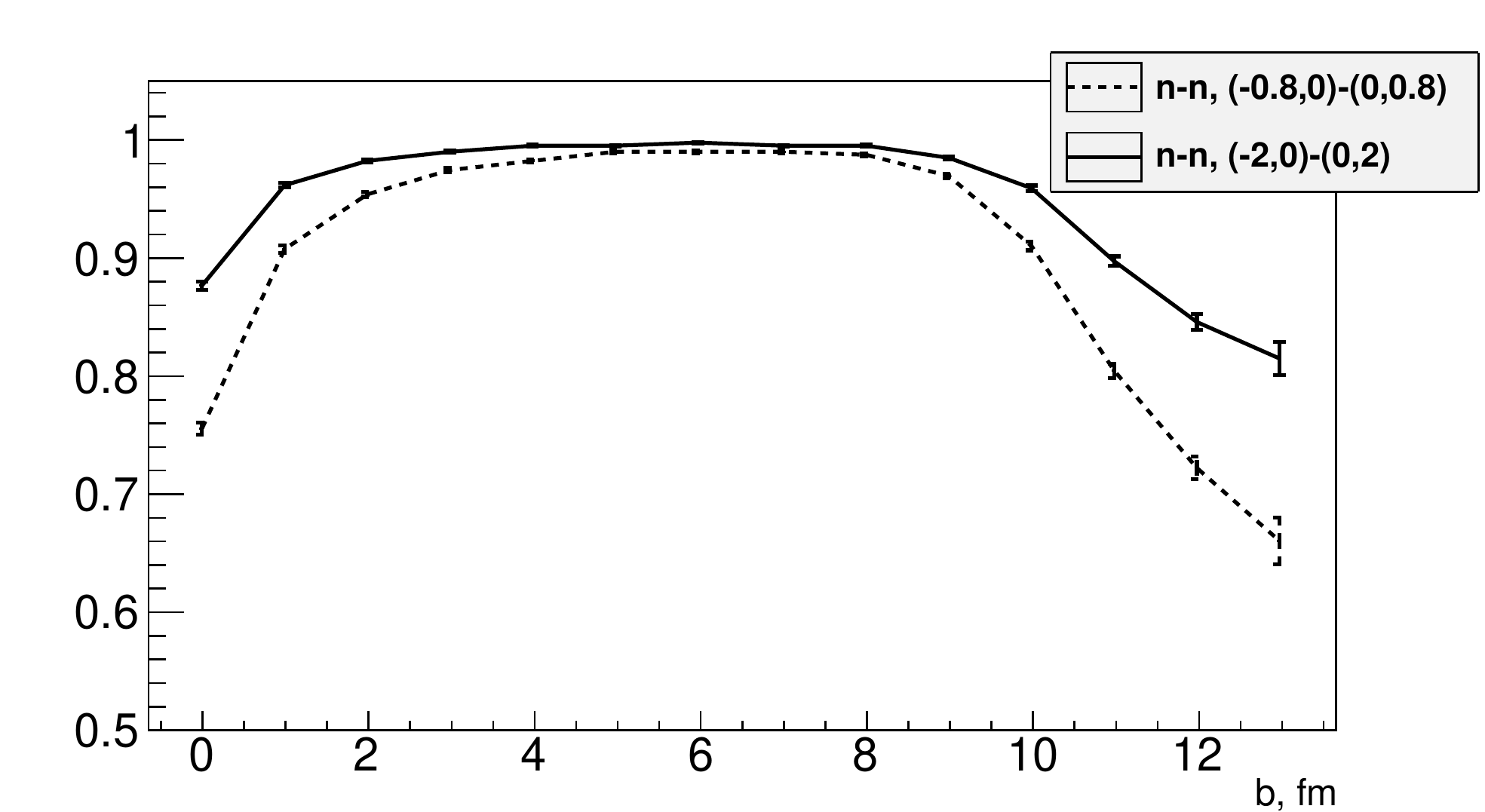}
\includegraphics[width=.7\textwidth]{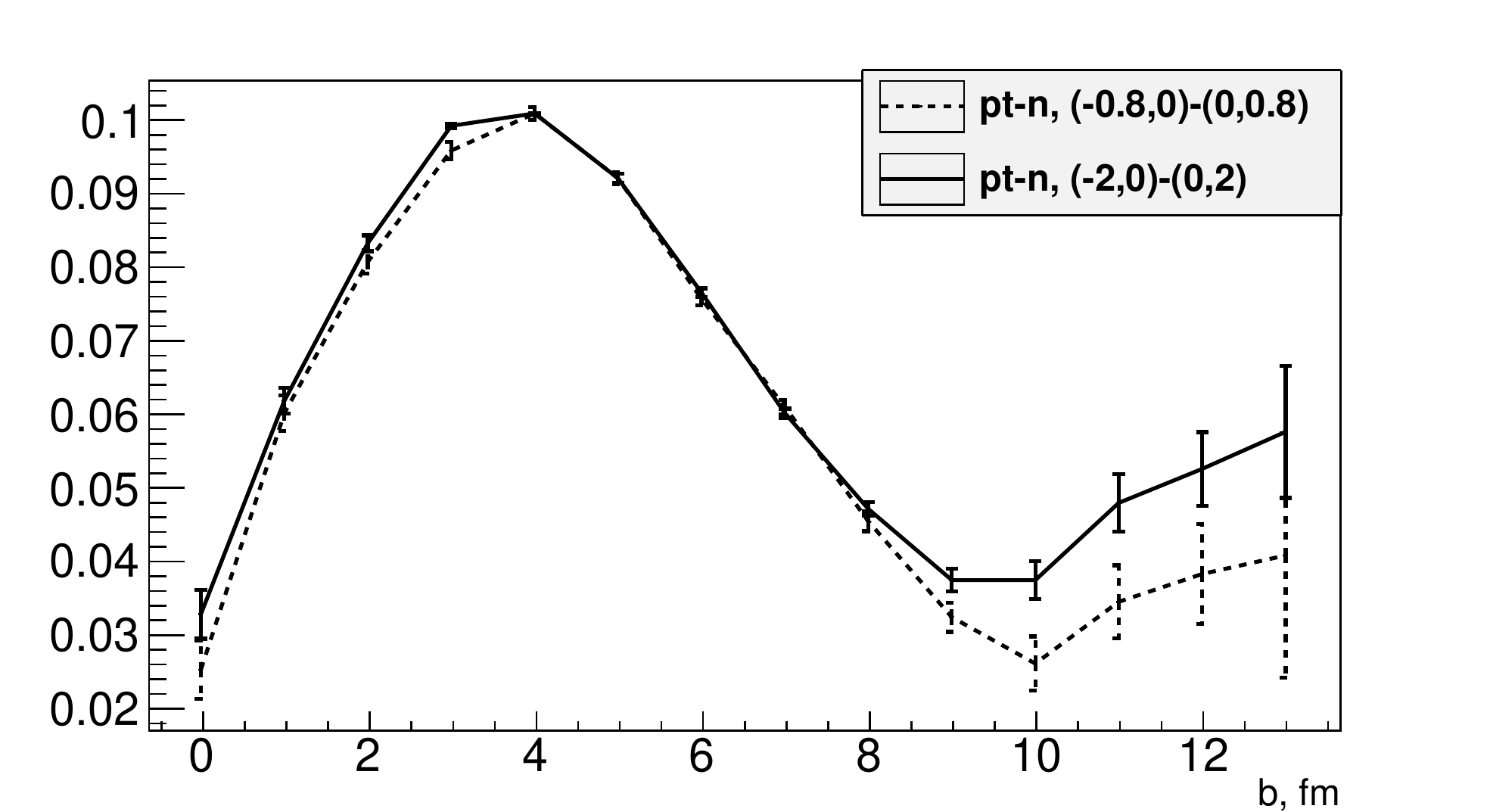}
\includegraphics[width=.7\textwidth]{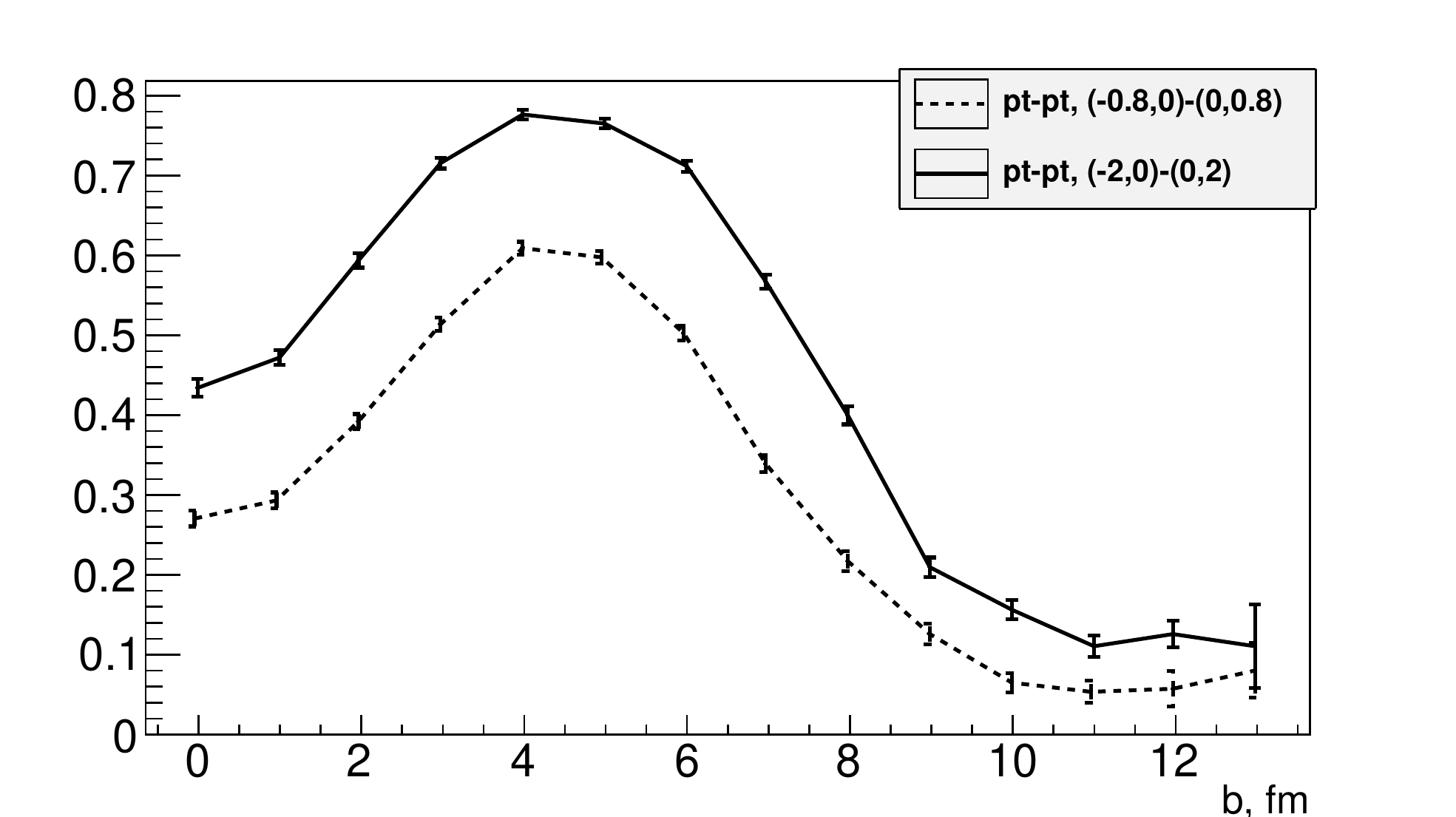}
\caption{N-n, pt-n and pt-pt correlation coefficients
with fixed impact parameter,
calculated in MC model for Pb-Pb collisions at $\sqrt{s_{_{NN}}}= 2.76$ TeV
}
\label{bAA}
\end{figure}

On fig. \ref{bAA} long-range correlation coefficients of $n-n$, $p_t-n$ and $p_t-p_t$ correlations are shown for PbPb collisions at the energy of LHC (2.76 TeV). Two different configurations of forward and backward rapidity windows were considered: windows in width 0.8 and 2 rapidity units.

It is noticeable, that even at the fixed impact parameter $n-n$ correlation are very strong. The correlation coefficient decreases in the central collisions that it expected due to string fusion effects (decrease of n-n correlation) are important in central PbPb collisions.
High value of n-n correlations in semi-central and peripheral collisions are due to volume fluctuation effects.

The large value of $p_t-p_t$ correlations is remarkable and dominates over $p_t-n$ correlations. This fact is in agreement with predictions \cite{VecherninYadPt} stated that at LHC energies $p_t-p_t$ correlation coefficient can reach the level of $n-n$ correlations.

\FloatBarrier

\section{Conclusion}

Long-range correlations are calculated in the framework of the string fusion model, taking into account finite rapidity length of the strings.

N-n, pt-n and pt-pt correlation functions were calculated and it was shown that they demonstrate non-linear behaviour.
The dependence of the correlation coefficients on the width of rapidity windows and the gap between them is calculated and compared with the predictions of the model of independent emitters.

The developed model enabled to describe also AA interactions without referring to the Glauber picture of nucleon-nucleon collisions.
The number of soft binary collisions in this approach is proved to be less than in the Glauber approach.

Correlation coefficients for Pb-Pb collisions are estimated in case of fixed impact parameter.
In Pb-Pb collisions at LHC energy the strength of pt-pt correlation is larger compared to pt-n correlation.

%we have to cite all of them \cite{06p1295, 88p191, Aamodt:2010cz, Aamodt:2010pp, Abelev:2009ag, Alexopoulos95p155, alicetech, Amelin:1993cs, Armesto07p201, Armesto:2000xh, Arneodo1985249, Asryan09p208, Brogueira:2009nj, Brombergetal70p25, Bzdak:2009xq, Dash12p1250079, DELPHICollaboration91p185, Dumitru08p91, Feofilov04p222, Khachatryan:2010us, Kharzeev:2004if, Konchakovski:2008cf, PhysRevD.37.3127, RAkers94p417, Tarnowsky:2008am, Uhlig78p15, V.V.Aivazyan89p533, Walker04p034007, WBraunschweig89p193, Wraight11p1}
% \cite{Wraight11p1}.
%Hello \cite{Wraight11p1, Armesto07p201,  %DELPHICollaboration91p185, Dumitru08p91} aaa.

\acknowledgments

The authors are grateful to M.A. Braun, G.A. Feofilov, T.A. Drozhzhova and I.G. Altsybeev for numerous useful discussions. This work was partially supported by the RFFI grant 12-02-00356-a.

\bibliography{papers}

\newpage
\appendix
\section*{Appendix}

\begin{figure}[!htbp]\center
\includegraphics[width=.6\textwidth]{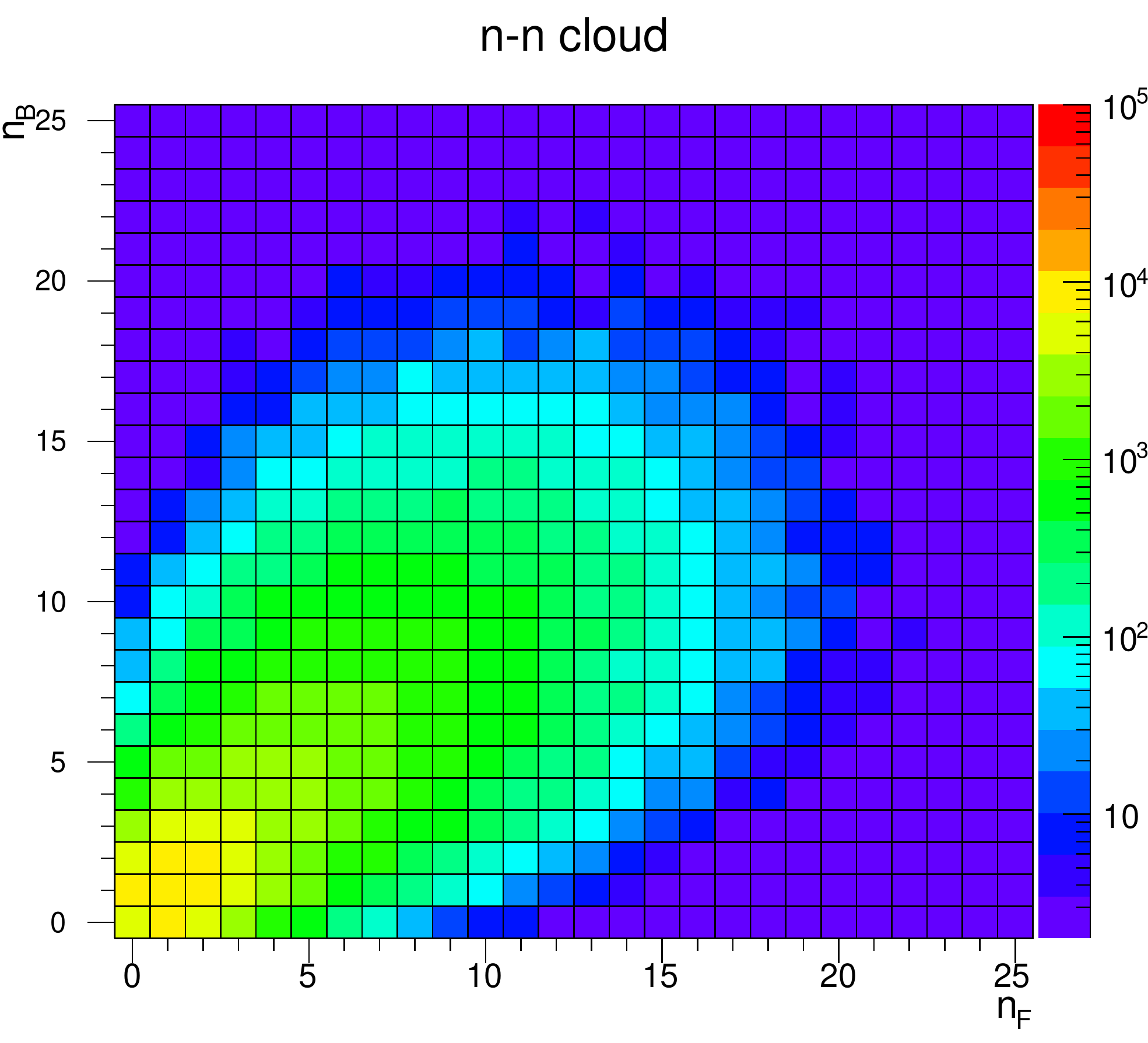}
\caption{2d n-n correlation histogram, calculated
in Monte Carlo model for pp interactions at 7TeV,
rapidity windows are (-0.8, 0), (0, 0.8).
}
\label{nncloud2}
\end{figure}

At fig. \ref{nncloud2} an example of
n-n correlation cloud is shown. The colour denotes
the number of events with given $n_F$ and $n_B$.
N-n correlation function is obtained by regression of this histogram.

\FloatBarrier

%\begin{thebibliography}{99}
%  \bibitem{...} ....
%\end{thebibliography}

\end{document}